\documentclass[twocolumn,trackchanges]{aastex701}
\usepackage{float}  
\usepackage{array}

\begin{document}

\title{Insights into the Exoplanet Radius Valley from Host-Star Ages, Activity, Chemistry, and Birth Radius}

\author[orcid=0000-0003-3957-9067]{Xunzhou Chen}
\affiliation{School of Science, Hangzhou Dianzi University, Hangzhou, PR China}
\affiliation{National Astronomical Data Center Zhijiang Branch, Hangzhou, PR China}
\email[show]{cxz@hdu.edu.cn}

\author[orcid=0000-0003-0795-4854]{Tiancheng Sun}
\affiliation{CAS Key Laboratory of Optical Astronomy, National Astronomical Observatories, Chinese Academy of Sciences, Beijing 100101, China}
\email[show]{suntc@bao.ac.cn}

\author[0000-0003-4769-3273]{Yuxi (Lucy) Lu}
\affiliation{Department of Astronomy, The Ohio State University, Columbus, 140 W 18th Ave, OH 43210, USA}
\affiliation{Center for Cosmology and Astroparticle Physics (CCAPP), The Ohio State University, 191 W. Woodruff Ave., Columbus, OH 43210, USA}
\email{lucylulu12311@gmail.com}

\author{Zixuan Lu}
\affiliation{School of Physics and Astronomy, Beijing Normal University, Beijing, China}
\email{202231160008@mail.bnu.edu.cn}

\author[orcid=0009-0009-1338-1045]{Lifei Ye}
\affiliation{School of Science, Xichang University, Xichang 615013, China}
\email{201831160010@mail.bnu.edu.cn}

\begin{abstract}
The radius valley—a bimodal feature in the size distribution of close-in small exoplanets—is widely interpreted as a signature of atmospheric loss and therefore provides a key constraint on the formation and atmospheric evolution of these planets. We investigate its dependence on host-star properties using 769 planets orbiting 558 stars, for which we derive stellar ages, chromospheric activity ($\log R^+_{\rm HK}$), and Galactic birth radius, together with elemental abundances. We find that the radius valley is not fully established at ages $\lesssim 3$~Gyr and evolves over gigayear timescales, with its prominence strongly affected by stellar population mixing. The dependence on magnetic activity is non-monotonic: a clear valley is present even among magnetically quiet stars, while highly active systems do not show a systematically stronger depletion. The valley morphology also varies with stellar composition: the valley is strongest in metal-poor stars, weakens near solar metallicity, and partially strengthens again at the highest metallicities. In addition, the valley shows sensitivity to refractory element ratios such as [Mg/Si], while correlations with [C/O] are weaker, indicating a dependence on planetary interior structure. Our results are more consistent with a dominant role for core-powered atmospheric mass loss than with purely irradiation-driven photoevaporation. Finally, the radius valley also depends on the Galactic birth environment, with systems near the estimated solar birth radius ($R_{\rm birth} \simeq 4.5 \pm 0.4$~kpc) showing a high fraction of Earth-like planets and a well-defined bimodal structure, suggesting that the Solar System formed in a region with a well-developed Earth-sized planet population.
\end{abstract}

\keywords{Exoplanets -- Stellar fundamental parameters -- Galactic Chemical Evolution}


\section{Introduction}
Since the first detection of a planet around a main-sequence star \citep{1995Natur.378..355M}, ground- and space-based surveys have established a census of over 5000 confirmed exoplanets (NASA Exoplanet Archive; \citet{2013PASP..125..989A}). This wealth of discoveries has enabled statistical studies of planetary populations and their formation pathways. Among them, the radius valley—a pronounced deficit of planets at radius of $\sim2.0\,R_\oplus$ separating super-Earths from sub-Neptunes—has emerged as one of the most robust population-level features in the Kepler sample (e.g., \citet{2017AJ....154..109F, 2018AJ....155...89P}). Although early measurements were limited by planetary-radius uncertainties comparable to the valley width, subsequent improvements based on asteroseismic constraints and Gaia parallaxes demonstrated that the valley persists and cannot be explained by observational scatter alone, establishing it as a real, though sparsely populated, feature of the exoplanet population \citep{2018MNRAS.479.4786V, 2018AJ....156..264F, 2020AJ....160...89P}.

The physical origin of this feature has been the subject of extensive theoretical investigation. Planet-evolution models incorporating photoevaporation predicted the emergence of a radius valley prior to its observational discovery \citep{2013ApJ...776....2L, 2013ApJ...775..105O, 2014ApJ...795...65J, 2016ApJ...831..180C}. Alternatively, atmospheric mass loss driven by the cooling luminosity of planetary cores—the so-called core-powered mass-loss mechanism—can likewise reproduce the observed bimodal radius distribution \citep{2018MNRAS.476..759G, 2019MNRAS.487...24G, 2020MNRAS.493..792G}. These two mechanisms constitute the leading theoretical explanations for the radius valley, and distinguishing between them requires understanding how the valley’s location and contrast depend on planetary and host-star properties. 

Motivated by this expectation, numerous studies have investigated how the morphology of the radius valley depends on stellar and planetary parameters. In the radius–period plane, the valley center exhibits a robust anti-correlation with orbital period or stellar irradiation \citep{2017AJ....154..109F,2018MNRAS.479.4786V,2019MNRAS.487.5062M,2019ApJ...875...29M,2020ApJ...890...23L}, supporting a dominant role for atmospheric mass loss in shaping close-in planets. Beyond orbital parameters, host-star properties further modulate the valley morphology, with its location reported to shift toward larger radius around more massive stars \citep{2018AJ....156..264F,2019ApJ...874...91W,2020AJ....159..211C,2021MNRAS.501.5309H,2021MNRAS.507.2154V}, while recent work further suggests that the valley depth itself depends on host-star mass, being significantly shallower around low-mass stars \citep{2024MNRAS.531.3698H}. In addition, stellar metallicity influences both the characteristic planet sizes and the width of the valley, with metal-rich stars preferentially hosting larger planets and exhibiting a broader valley \citep{2018PNAS..115..266D,2018AJ....155...89P,2018MNRAS.480.2206O}. Stellar age and magnetic activity further probe the timescale and efficiency of atmospheric escape, and observations indicate that the relative occurrence of super-Earths and sub-Neptunes evolves with stellar age, with younger systems displaying a wider and more depleted valley \citep{David_2021,2021ApJ...911..117S,2022AJ....163..249C}. Moreover, because stellar age, metallicity, and activity are themselves shaped by the chemo-dynamical evolution of the Milky Way, these trends raise the possibility that the morphology of the radius valley also carries imprints of Galactic environment and stellar birth history.

Advancing such studies critically depends on the precision and accuracy with which host-star properties can be measured. Recent progress in astrometry, spectroscopy, and data-driven modeling has substantially improved constraints on stellar ages, chemical abundances, and activity indicators. In particular, the combination of exquisite Gaia DR3 astrometry \citep{2023gaia} with large-scale spectroscopic surveys such as LAMOST \citep{2012RAA....12.1197C}, GALAH \citep{2015MNRAS.449.2604D}, and APOGEE \citep{2017AJ....154...94M} now enables homogeneous and precise characterization of host-star properties across large samples, providing a robust foundation for investigating population-level trends in exoplanet demographics. In this work, we combine Gaia DR3 with LAMOST DR9 DD-Payne stellar parameters and abundances \citep{2025ApJS..279....5Z} to construct a sample of 769 planets orbiting 558 host stars, with homogeneous measurements of stellar atmospheric parameters and detailed chemical compositions. Building on these data, we derive a consistent set of stellar ages, the chromospheric activity indicator $R^{+}_{\rm HK}$, and the stellar birth radius $R_{\rm birth}$. We then quantify how the morphology of the radius valley varies across this multi-dimensional host-star parameter space, and assess how planetary evolution and Galactic environment jointly shape the observed demographics of planets. This paper is organized as follows. Section~\ref{sec:data} describes the data, sample selection, and methodology. The main results are presented in Section~\ref{sec:result}, and our conclusions are summarized in Section~\ref{sec:con}.

\section{Data and Method} \label{sec:data}

\subsection{Sample Selection}
We adopt the stellar sample of 27,626 dwarfs and subgiants compiled by \citet{2025ApJ...995...33C} (hereafter C25), constructed from a cross-match of Kepler stars \citep{2020AJ....159..280B} with LAMOST DR9 DD-Payne \citep{2025ApJS..279....5Z} and Gaia DR3 \citep{2023gaia}, followed by quality cuts. The sample includes precise measurements of $T_{\rm eff}$, $\log g$, [Fe/H], and 22 individual elemental abundances from LAMOST DD-Payne \citep{https://doi.org/10.5281/zenodo.15223381}, with [Fe/H] corrected for non-LTE effects, $T_{\rm eff}$ calibrated to the IRFM scale, and $\log g$ validated against asteroseismology. Parallaxes, proper motions, and radial velocities are taken from Gaia DR3. Planetary properties are adopted from the Kepler DR25 catalog \citep{Thompson_2018, k2pandc}. Applying cuts of $P_{\rm orb} < 100\ \mathrm{days}$ and $0.5 < R_p/R_\oplus < 4$ to ensure detection completeness and to isolate the radius-valley regime, we obtain a sample of 769 planets orbiting 558 host stars. The construction of this parent sample is summarized in Table~\ref{tab:stellar_selection} (steps~1--8). We further derive additional stellar parameters and define parallel subsamples for different analyses: 520 planets around 381 stars with age and mass estimates, 651 planets around 472 stars with measured $\log R^{+}_{\rm HK}$, and 508 planets around 372 stars with inferred stellar birth radius. These subsamples are summarized in Table~\ref{tab:stellar_selection} (steps~9--11), and the derivation methods are described below. 

Because different stellar parameters are available only for subsets of the parent sample after applying quality cuts, the number of planets included in each analysis differs. Each parameter is therefore analyzed using the largest available subsample rather than restricting the analysis to the smallest common subset. This approach maximizes the statistical significance of the radius-valley measurements while preserving the relative planet-number ratios within each subsample.

\begin{table*}
\centering
\scriptsize
\caption{Summary of the Sample Selection}
\label{tab:stellar_selection}
\begin{tabular}{clc}
\hline
Step & Selection Criteria & Number of Stars Remaining \\
\hline
1 & Kepler stellar sample \citep{2020AJ....159..280B} & 186,301 \\
2 & Cross-match with LAMOST DR9 DD-Payne ($<1.5^{\prime\prime}$) \citep{2025ApJS..279....5Z} & 76,058 \\
3 & Cross-match with Gaia DR3 \citep{2023gaia} & 74,581 \\
4 & SNRG $\geq 20$, $\chi^2$ flags $\leq 3$, valid abundance flags, RUWE $\leq 1.2$, available CDPP, and $5000 \leq T_{\rm eff} \leq 6800$~K & 33,219 \\
6 & Initial stellar mass cut: $M < 1.4\,M_\odot$ & 27,626 \\
7 & Cross-match with Kepler DR25 planets with $P_{\rm orb} < 100$~days and $R_p > 0.5\,R_\oplus$ & 601 (830 planets) \\
8 & Selection of planets with $R_p < 4\,R_\oplus$ & 558 (769 planets) \\
\hline
9  & \textit{From Step~8:} stars with reliable ages ($\rm age < 15$~Gyr, $\rm age\!-\!2\sigma < 13.8$~Gyr, relative error $<100\%$) & 381 (520 planets) \\
10 & \textit{From Step~8:} stars with derived $\log R^{+}_{\rm HK}$ & 472 (651 planets) \\
11 & \textit{From Step~8:} stars with derived $R_{birth}$ & 372 (508 planets) \\
\hline
\end{tabular}
\tablecomments{
Steps~1--7 follow the stellar sample selection of \citet{2025ApJ...995...33C}. 
Step~8 applies an additional planet-radius cut ($R_p<4\,R_\oplus$). 
Steps~9--11 define three \emph{parallel} subsamples drawn from the Step~8 sample (rather than sequential cuts) for analyses requiring stellar ages, $\log R^{+}_{\rm HK}$, and stellar birth radius, respectively. 
}
\end{table*}

\subsection{Stellar Luminosity and Radius}
\begin{figure}[htbp]
    \centering
    \includegraphics[width=0.4\textwidth]{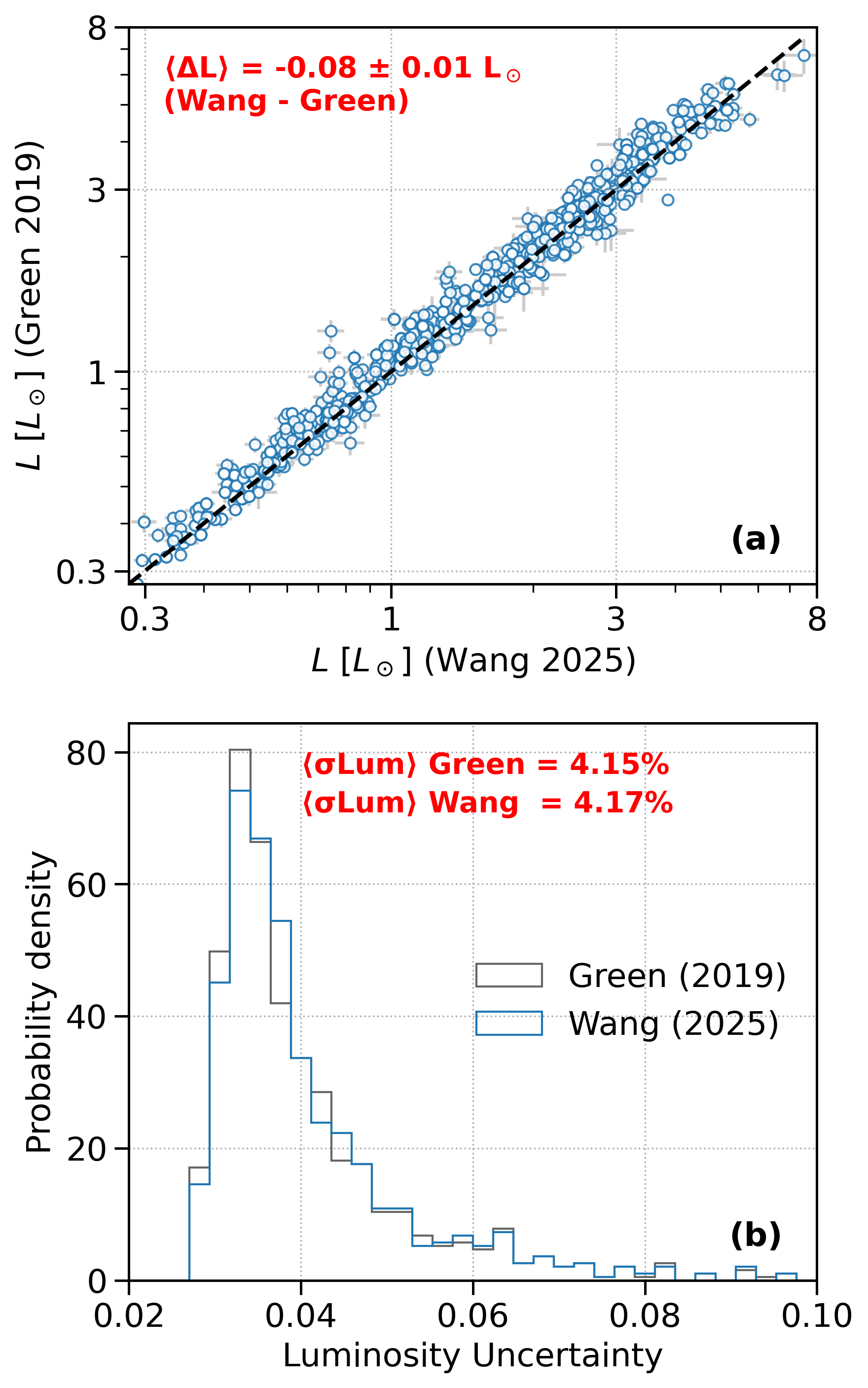}
    \caption{
    Comparison of stellar luminosities derived using different extinction maps. 
    \textbf{(a)} One-to-one comparison of luminosities in logarithmic scale. 
    \textbf{(b)} Probability density distributions of relative luminosity uncertainties.
    }
    \label{fig:luminosity_comparison}
\end{figure}

In C25, stellar luminosities were derived using a direct method implemented in \texttt{isoclassify}, combining 2MASS $K$-band apparent magnitudes, Gaia DR3 parallaxes corrected for the global zero-point offset following \citet{2021A&A...649A...4L}, and extinction values from the dust map of \citet{2019ApJ...887...93G}. In this work, we recompute stellar luminosities using extinction corrections from \citet{Wang_2025}, derived from LAMOST spectroscopy combined with Gaia XP low-resolution spectra. Its high angular resolution (3$'$--6$'$) allows accurate $E(B-V)$ estimates in low-extinction, high-latitude regions. Moreover, the extinction map is based on LAMOST spectra, which is particularly well suited to our sample. Figure~\ref{fig:luminosity_comparison} compares the stellar luminosities derived using this updated method with previous values. Panel (a) presents a one-to-one comparison in logarithmic scale, showing that the newly derived luminosities are systematically lower by $\sim 0.08~L_\odot$ on average. Panel (b) shows the probability density distributions of relative luminosity uncertainties for both datasets, which are broadly consistent, with an average uncertainty of $\sim 4.1\%$, indicating comparable precision between the two measurements. Stellar radius are then derived from the input effective temperatures and the inferred stellar luminosities using the Stefan--Boltzmann relation,
\begin{equation}
R = \left(\frac{L}{4\pi\sigma T_{\rm eff}^4}\right)^{1/2},
\end{equation}
where $L$ is the stellar luminosity, $T_{\rm eff}$ is the effective temperature, and $\sigma$ is the Stefan--Boltzmann constant.

\subsection{Stellar Ages}
Based on the updated stellar luminosities, we re-derived stellar ages using the same Bayesian isochrone fitting framework adopted in C25. The method follows the Bayesian formalism of \citet{2010A&A...522A...1K} and \citet{2010ApJ...710.1596B}, in which the likelihood is constructed from the observed stellar parameters $(T_{\rm eff}, L, \mathrm{[Fe/H]})$ and compared against grids of stellar evolutionary models. 
For each star, the posterior probability of a model $M_i$ given the data $D$ is computed via Bayes' theorem:
\begin{equation}
p(M_i|D) \propto p(D|M_i) \, p(M_i),
\end{equation}
where $p(M_i)$ is the prior probability (assumed uniform) and $p(D|M_i)$ is the likelihood function, which we approximate as
\begin{equation}
p(D|M_i) = \prod_{j} \frac{1}{\sqrt{2\pi}\sigma_j} \exp\left[-\frac{(O_{j,{\rm obs}} - O_{j,{\rm model}})^2}{2\sigma_j^2}\right],
\end{equation}
with $O_j$ representing the observed stellar parameters and $\sigma_j$ their associated uncertainties. Stellar ages and uncertainties are then estimated from the posterior probability distribution, typically adopting the mean and standard deviation. We employ the $\alpha$-enhanced stellar models of \citet{2023MNRAS.523.1199S}, selecting for each star the subset of models closest to its measured $[\alpha/\mathrm{Fe}]$, calculated as the error-weighted mean of Mg, Si, Ca, and Ti abundances from the LAMOST DR9 DD-Payne catalog, where these elemental abundances are derived under the LTE assumption.

\begin{figure}[htbp]
    \centering
    \includegraphics[width=0.4\textwidth]{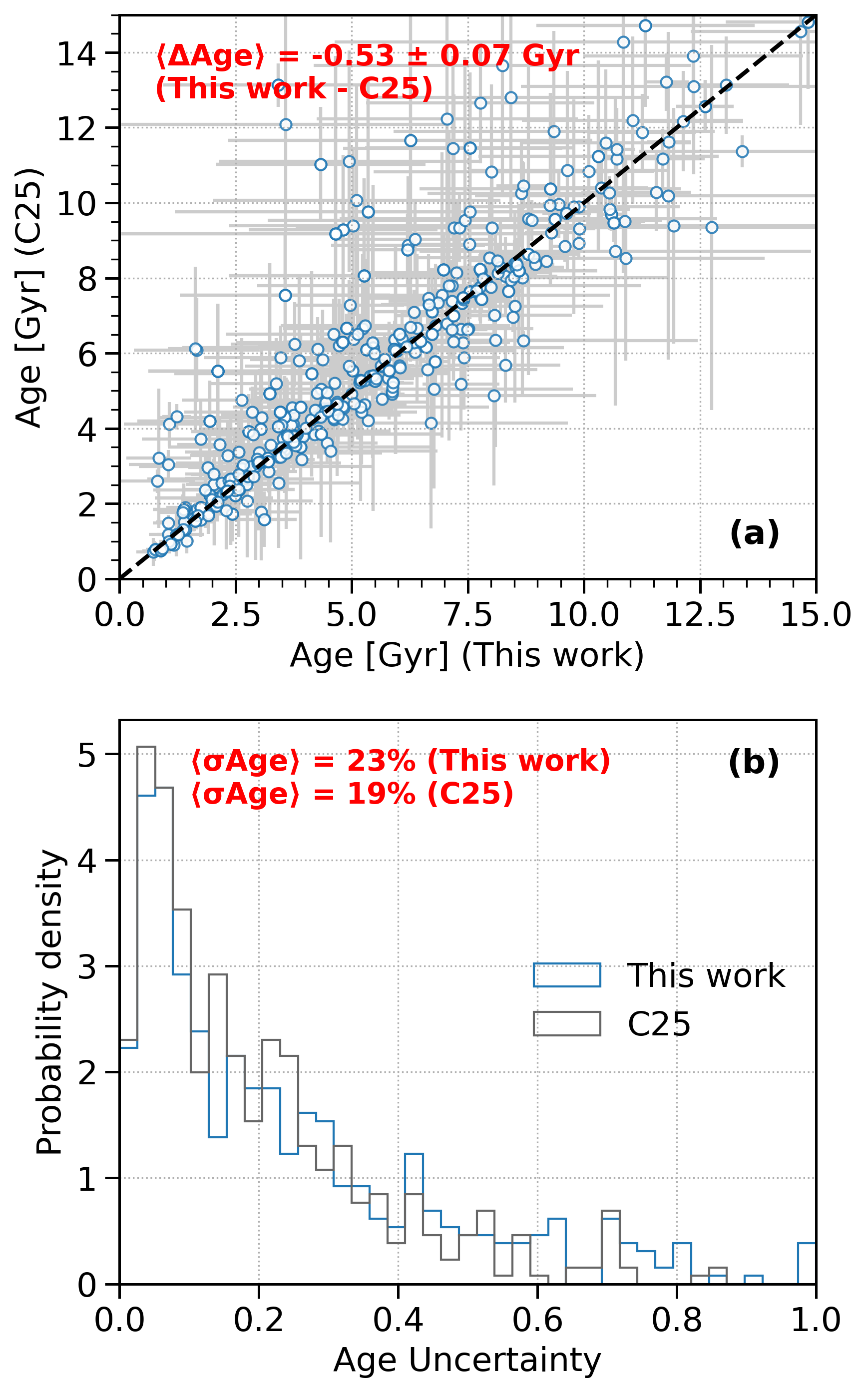}
    \caption{
    Comparison of stellar ages derived in this work and in C25. 
    \textbf{(a)} One-to-one comparison of stellar ages. The dashed line indicates equality. The average age difference is $\sim$0.5~Gyr, with our ages being younger. 
    \textbf{(b)} Probability density distributions of relative age uncertainties, showing overall consistency, with most stars having uncertainties below 40\% and a peak near 10\%. 
    }
    \label{fig:age_comparison}
\end{figure}

Figure~\ref{fig:age_comparison} compares the newly derived stellar ages with those from C25. Stars with reliable ages ($\rm age - 2\sigma < 13.8~Gyr$, relative uncertainty $< 100\%$, age $< 15~\rm Gyr$) are selected. Panel~(a) presents a one-to-one comparison, showing that the majority of stars lie close to the one-to-one line, indicating agreement between the two results. The mean age difference, $\langle \Delta \mathrm{Age} \rangle = \mathrm{Age}_{\rm new} - \mathrm{Age}_{\rm C25}$, is approximately $-0.5$~Gyr, indicating that the updated ages are systematically younger. Panel~(b) shows the distributions of relative age uncertainties, which are broadly consistent between the two datasets. Most stars exhibit relative errors below 40\%, peaking near 10\%, with the mean relative uncertainty for the updated ages being $\sim 23\%$, comparable to $\sim 19\%$ in C25, with the slightly larger uncertainty reflecting the shift toward younger ages, where isochrone constraints are generally weaker. This demonstrates that the updated sample maintains high-precision age estimates.

\subsection{Stellar Activity}

The $R^{+}_{\rm HK}$ is a widely used proxy of stellar chromospheric activity \citep[see definition in][]{2013A&A...549A.117M}. Unlike the classical $R'_{\rm HK}$ index, $R^{+}_{\rm HK}$ further subtracts the basal chromospheric flux and therefore isolates the purely magnetic activity-related Ca~II H\&K emission. As a result, $R^{+}_{\rm HK}$ provides a cleaner measure of stellar magnetic activity, particularly for low-activity stars. For each target, we measured the activity indices from the LAMOST low-resolution (LRS) DR12 spectra following the methodology described in \citet{Yu2024MNRAS.530.2953Y}.
 
Following the classical Mount Wilson definition 
\citep{wilson1968ApJ...153..221W, vaughan1978PASP...90..267V, duncan1991ApJS...76..383D, baliunas1995ApJ...438..269B}, 
the instrumental $S$-index is computed as \citep{lovis2011arXiv1107.5325L,karoff2016NatCo...711058K}:
\begin{equation}\label{eq:shk}
S_{L} = 8\alpha\frac{H+K}{R+V},
\end{equation}
where $\alpha$ is a calibration factor set to 1.8. 
The $R$ and $V$ terms correspond to 20~\AA\ rectangular continuum bandpasses spanning 
3992.20--4012.20~\AA\ and 3892.17--3912.17~\AA, respectively. 
The $H$ and $K$ terms are measured using triangular bandpasses centered at 
3969.59~\AA\ and 3934.78~\AA, with a full width at half maximum (FWHM) of 1.09~\AA.

For stars with two or more visits, we report a signal-to-noise ratio (S/N) weighted mean $S$-index,
\begin{equation}\label{eq:shk_weight}
S = \frac{\sum_{i=1}^{n} S_{L,i}\,\mathrm{SNR}_{i}}{\sum_{i=1}^{n} \mathrm{SNR}_{i}},
\end{equation}
where $n$ is the number of available spectra and $\mathrm{SNR}_{i}$ is the signal-to-noise ratio of the $i$th spectrum.

To place the activity metric on a bolometric, photosphere-subtracted scale, 
we convert the $S$-index to $R^{+}_{\rm HK}$ following \citet{mittag2013}:
\begin{equation}\label{eq:rhkplus}
R^{+}_{\rm HK} =
\frac{\mathcal{F}_{\rm HK} - \mathcal{F}_{\rm HK,phot} - \mathcal{F}_{\rm HK,basal}}
{\sigma T_{\rm eff}^{4}},
\end{equation}
where $\sigma$ is the Stefan--Boltzmann constant, 
$\mathcal{F}_{\rm HK,phot}$ represents the photospheric contribution to the Ca~II H\&K lines \citep{Noyes1984ApJ...279..763N}, 
and $\mathcal{F}_{\rm HK,basal}$ is the basal chromospheric flux \citep{Schrijver_RHKplus_1987A&A...172..111S}. The surface flux proxy $\mathcal{F}_{\rm HK}$ is derived from the $S$-index following \citet{1982A&A...107...31M}:
\begin{equation}\label{eq:fhk}
\mathcal{F}_{\rm HK} = 10^{8.25 - 1.67(B-V)}\, S,
\end{equation}
where $B-V$ is the stellar color index. 
Both $\mathcal{F}_{\rm HK,phot}$ and $\mathcal{F}_{\rm HK,basal}$ are estimated from empirical relations as functions of $B-V$.
Uncertainties in $R^{+}_{\rm HK}$ are obtained by propagating the measurement uncertainties of the $S$-index.

\subsection{Stellar Birth Radius}

The stellar birth radius, $R_{\rm birth}$, is estimated following the framework developed by \citet{2024MNRAS.535..392L}, which extends the chemical--dynamical approach originally proposed by \citet{2018MNRAS.481.1645M}. Based on two independent suites of cosmological simulations, \citet{2024MNRAS.535..392L} showed that, at a given lookback time $\tau$, the radial metallicity gradient of the interstellar medium (ISM), $\nabla{\rm [Fe/H]}(\tau)$, is approximately linearly related to the metallicity spread of coeval stellar populations, quantified as $\mathrm{Range}\,\widetilde{\rm [Fe/H]}(\mathrm{age})$.

Assuming a radially linear ISM metallicity profile at the epoch of star formation, the birth radius of an individual star can be expressed as
\begin{equation}
R_{\rm birth}(\mathrm{age}, {\rm [Fe/H]}) =
\frac{{\rm [Fe/H]} - {\rm [Fe/H]}(R=0, \tau)}
{\nabla{\rm [Fe/H]}(\tau)} ,
\label{eqn:rb}
\end{equation}
where ${\rm [Fe/H]}(R=0, \tau)$ denotes the extrapolated central ISM metallicity at lookback time $\tau$, and $\nabla{\rm [Fe/H]}(\tau)$ is the corresponding radial metallicity gradient. Both quantities are obtained via interpolation of Table~A1 in \citet{2024MNRAS.535..392L}.

\subsection{Planet Radius}
Accurate planetary radius are crucial for reliably characterizing the radius valley, as the typical uncertainties in catalog-provided radius are comparable to the width of the valley. Such large uncertainties make it difficult to resolve the radius valley. To improve the precision, we recomputed planetary radius using stellar radius derived from our updated luminosities combined with the measured transit depths, following the relation given in Eq.~(3) of \citet{Wanderley_2025}. The transit depth, $\Delta F$, is defined as the ratio of the projected areas of the planet and the host star. The planetary radius can then be calculated as:
\begin{equation}
R_{\rm p} = \sqrt{\Delta F} \times R_\star,
\end{equation}
where $R_\star$ is the stellar radius. Transit depth values for the studied sample were taken from the Kepler DR25 catalog \citep{Thompson_2018, k2pandc}. The uncertainties of the planetary radius were estimated by propagating the errors in both the transit depth and the stellar radius. 

\begin{figure}[htbp]
    \centering
    \includegraphics[width=0.4\textwidth]{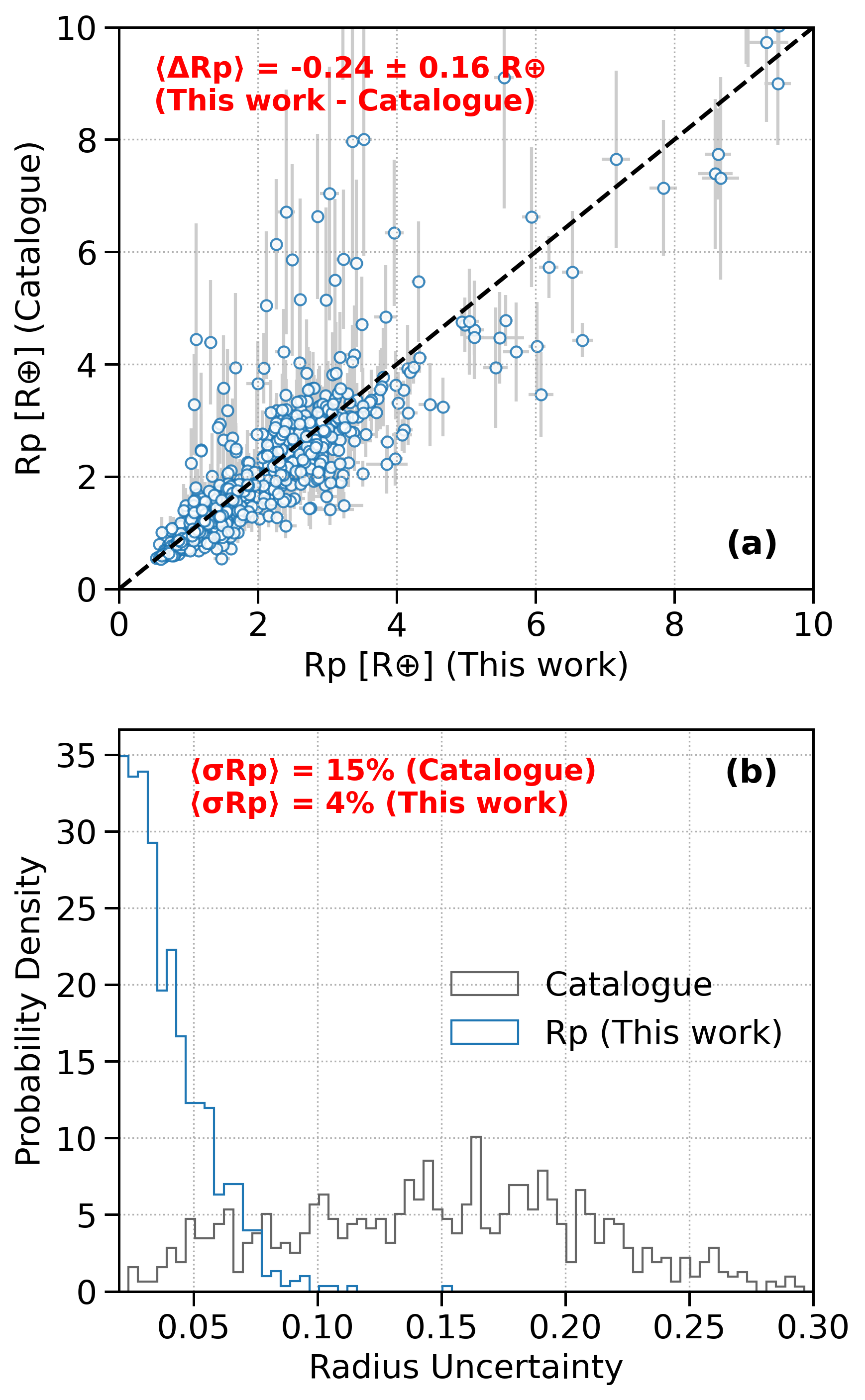}
    \caption{
    Comparison of planetary radius derived in this work and from the Kepler DR25 catalogue. \textbf{(a)} One-to-one comparison of planetary radius. The dashed line indicates equality. \textbf{(b)} Probability density distributions of relative radius uncertainties.
    }
    \label{fig:radius_comparison}
\end{figure}

As shown in Figure~\ref{fig:radius_comparison}, Panel~(a) presents a direct comparison between planetary radius derived in this work and those reported in the Kepler DR25 catalogue. Overall, our radius estimates are broadly consistent with the catalogue values and are, on average, smaller by about 0.24~$R_\oplus$, with a small subset of planets (our $R_p<4~R_\oplus$) having noticeably larger catalogue radius. Panel~(b) displays the distributions of relative radius uncertainties. Our derived radius typically have uncertainties below 10\%, with a mean of $\sim$4\%, significantly smaller than the catalogue values (mean $\sim$15\%), enabling a reliable characterization of the morphology of the radius valley.

\begin{figure}[htbp]
    \centering
    \includegraphics[width=0.4\textwidth]{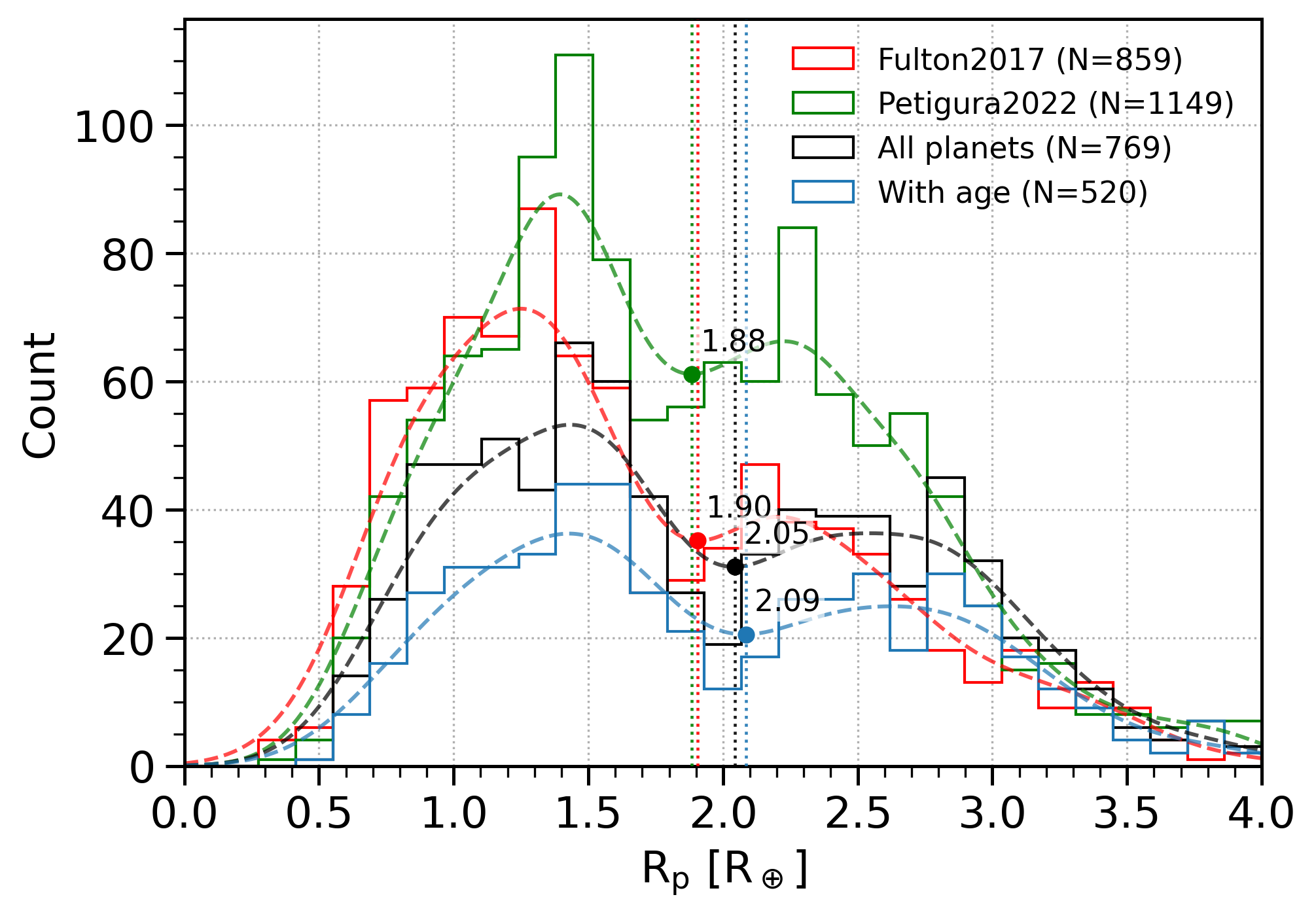}
    \caption{Comparison of planetary radius distributions for our sample and previous studies. All samples are restricted to $R_p < 4\,R_\oplus$. Histograms show the counts of planets in each bin, while dashed lines indicate KDE-based smoothed distributions.}
    \label{fig:radius_distribution}
\end{figure}

Figure~\ref{fig:radius_distribution} compares the planetary radius distribution with those reported in \citet{2017AJ....154..109F} and \citet{2022AJ....163..179P}. Our sample includes 769 planets in total, of which 520 have reliable host star age estimates. All four samples clearly exhibit a pronounced deficit of planets at $R_p \sim 2\,R_\oplus$, corresponding to the well-known radius valley. In each case, the peak of the radius distribution below the valley (super-Earths) is higher than that above it (sub-Neptunes).
Compared to literature samples, the location of the radius valley in our distribution is shifted toward larger radius by $\sim 0.1\ R_\oplus$, likely due to differences in sample selection and parameter estimation.

\subsection{Characterizing the Radius Valley Morphology}
\subsubsection{Fitting to the Radius Valley}

To characterize the radius valley as a function of stellar properties, we adopt a non-parametric approach following the methodology of \citet{2023MNRAS.519.4056H,2024MNRAS.531.3698H}. For a given planetary subsample, we first estimate the two-dimensional density in the radius--period plane using a kernel density estimator (KDE). We then apply a Gaussian Mixture Model (GMM) to separate the planets into two populations corresponding to the upper and lower sides of the valley. A linear boundary between the two populations is subsequently fitted using a support vector machine (SVM; \citealt{Vapnik2000}) with a linear kernel. 

While the radius valley is often parameterized as a power law in the $\log R_p$--$\log P_{\rm orb}$ plane, we find that this form does not effectively separate the two planet populations in our sample, likely because the restricted radius range ($R_p<4\,R_\oplus$) compresses the distribution in $\log R_p$ space and weakens the contrast between the two populations. We therefore adopt a semi-log parameterization, fitting the valley as a line in the $R_p$--$\log P_{\rm orb}$ plane.

\begin{equation}
R_p = A \log P_{\rm orb} + C,
\end{equation}
where $R_p$ is the planetary radius, $P_{\rm orb}$ is the orbital period, and $A$ and $C$ are the slope and intercept. To estimate uncertainties, we bootstrap-resample the planets within each subsample and refit the SVM for each realization. We report the mean $A$ and $C$ over all bootstrap iterations, and derive the $1\sigma$ confidence region from the 16th and 84th percentiles.

To further characterize the valley width, we define a valley region around the SVM boundary by adopting a fixed width of $0.4\,R_\oplus$ in planetary radius, motivated by previous studies that place the radius valley near $R_p \simeq 1.9 \pm 0.2\,R_\oplus$ \citep{2013ApJ...775..105O,2017AJ....154..109F,2022AJ....163..249C}:
\begin{equation}
R_p^{\rm upper/lower} = A \log P_{\rm orb} + (C \pm 0.2).
\end{equation}
Planets located within this valley region are considered representative of the radius valley, while those below correspond to the Earth-like population and those above to the sub-Neptune population. 

\begin{figure}[ht!]
\centering
\includegraphics[width=0.4\textwidth]{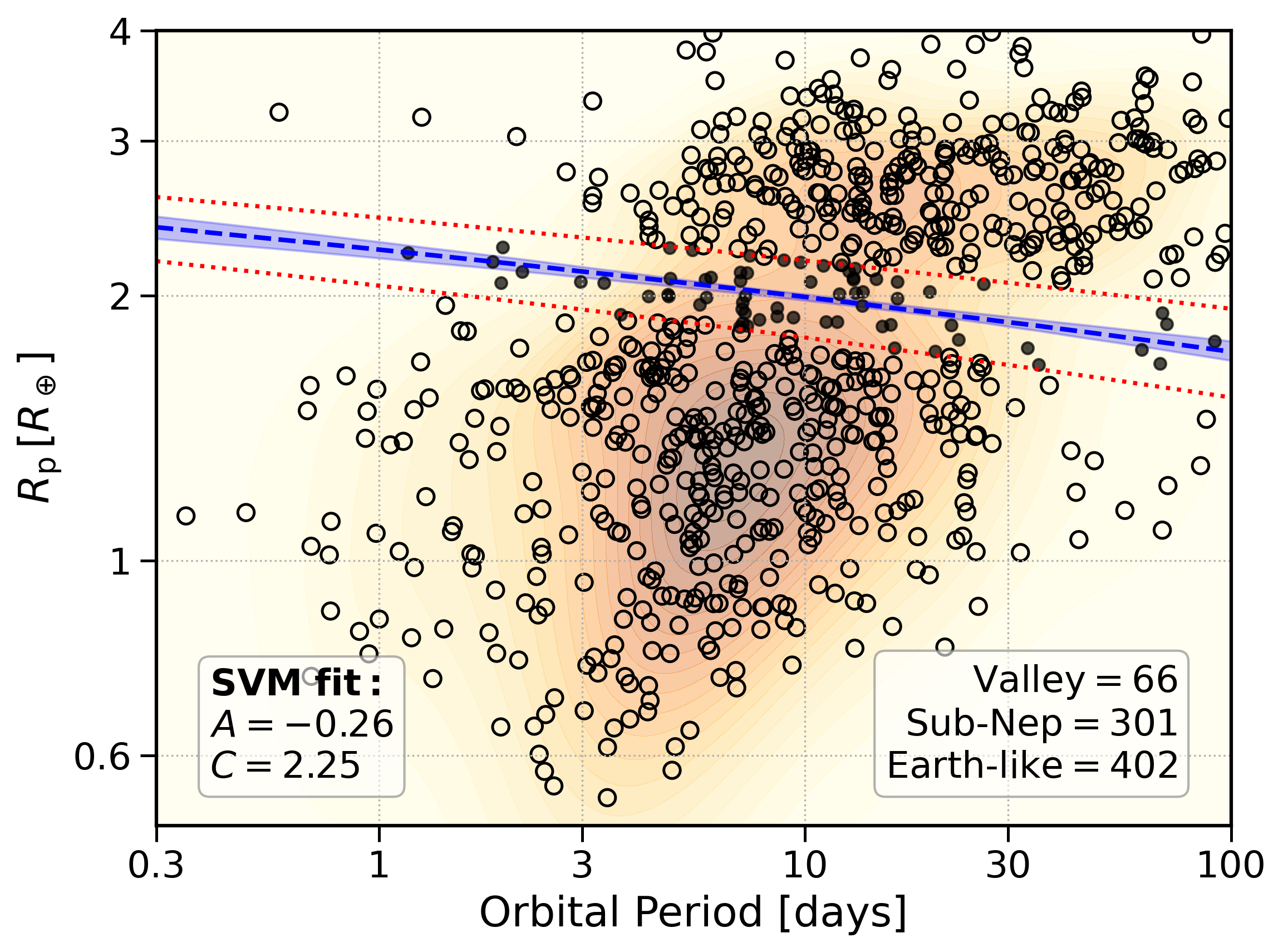}
\caption{
Planet radius-period distribution for the full sample. The blue dashed line shows the support vector machine (SVM) fit to the radius valley, and the shaded blue region indicates the $1\sigma$ uncertainty from bootstrap resampling. Red dotted lines mark $\pm0.2\,R_\oplus$ around the fit, defining the valley region used to classify planets within the valley. Contours represent the two-dimensional density estimated via a kernel density estimator (KDE), highlighting regions where planets are more densely populated.
}
\label{fig:rpgap_all}
\end{figure}

Applying this classification to the full sample, we examine the distribution of planets in the radius–period plane as shown in Figure~\ref{fig:rpgap_all}. The SVM fit to the valley yields $A=-0.26$ and $C=2.25$, implying that the valley center decreases with increasing orbital period: $R_p \approx 2.25\,R_\oplus$ at $P_{\rm orb}=1$~day and $R_p \approx 1.73\,R_\oplus$ at $P_{\rm orb}=100$~days. Note that this slope differs from previous measurements \citep{2018MNRAS.479.4786V,2019ApJ...875...29M} because we fit a linear relation in the $R_p$--$\log P_{\rm orb}$ plane instead of a power law in the $\log R_p$--$\log P_{\rm orb}$ plane. The contours also reveal the presence of two distinct planetary populations, which are well separated by the radius valley. Within the defined valley region, 66 planets are located in the valley, while 402 and 301 planets fall below (Earth-like) and above (Sub-Neptunes), respectively. Although the radius valley is clearly discernible, it contains planets and is not solely an artifact of measurement noise \citep{2018AJ....155...89P,2018MNRAS.479.4786V}.

\subsubsection{Parameterization of the Radius Valley}

Following a similar approach to \citet{2022AJ....163..249C}, we quantify the morphology of the radius valley using the relative number counts of planets in three regions of the radius--period plane: below the valley (``Earth-like''), above the valley (``Sub-Neptune''), and within the valley. For each subsample, planets are classified into these regions using a fixed-width band centered on the empirically identified valley locus, yielding the counts $N_{\rm Earth-like}$, $N_{\rm Sub-Nep}$, and $N_{\rm Valley}$.

We then define
\begin{equation}
A_{\rm valley} = \frac{N_{\rm Earth-like}}{N_{\rm Sub-Nep}},\ C_{\rm valley} = \frac{N_{\rm Earth-like}+N_{\rm Sub-Nep}}{N_{\rm Valley}}
\end{equation}

The parameter $A_{\rm valley}$ traces the relative dominance of Earth-like planets versus Sub-Neptunes on either side of the valley, while $C_{\rm valley}$ characterizes the depth of the valley by comparing the number of planets outside the valley to those within it. Uncertainties in both parameters are estimated via bootstrap resampling, with the 16th and 84th percentiles of the resulting distributions adopted as the $1\sigma$ confidence intervals.

\section{Results} \label{sec:result}
\subsection{Age Dependence of the Radius Valley}

\begin{figure*}
\centering
\includegraphics[width=\textwidth]{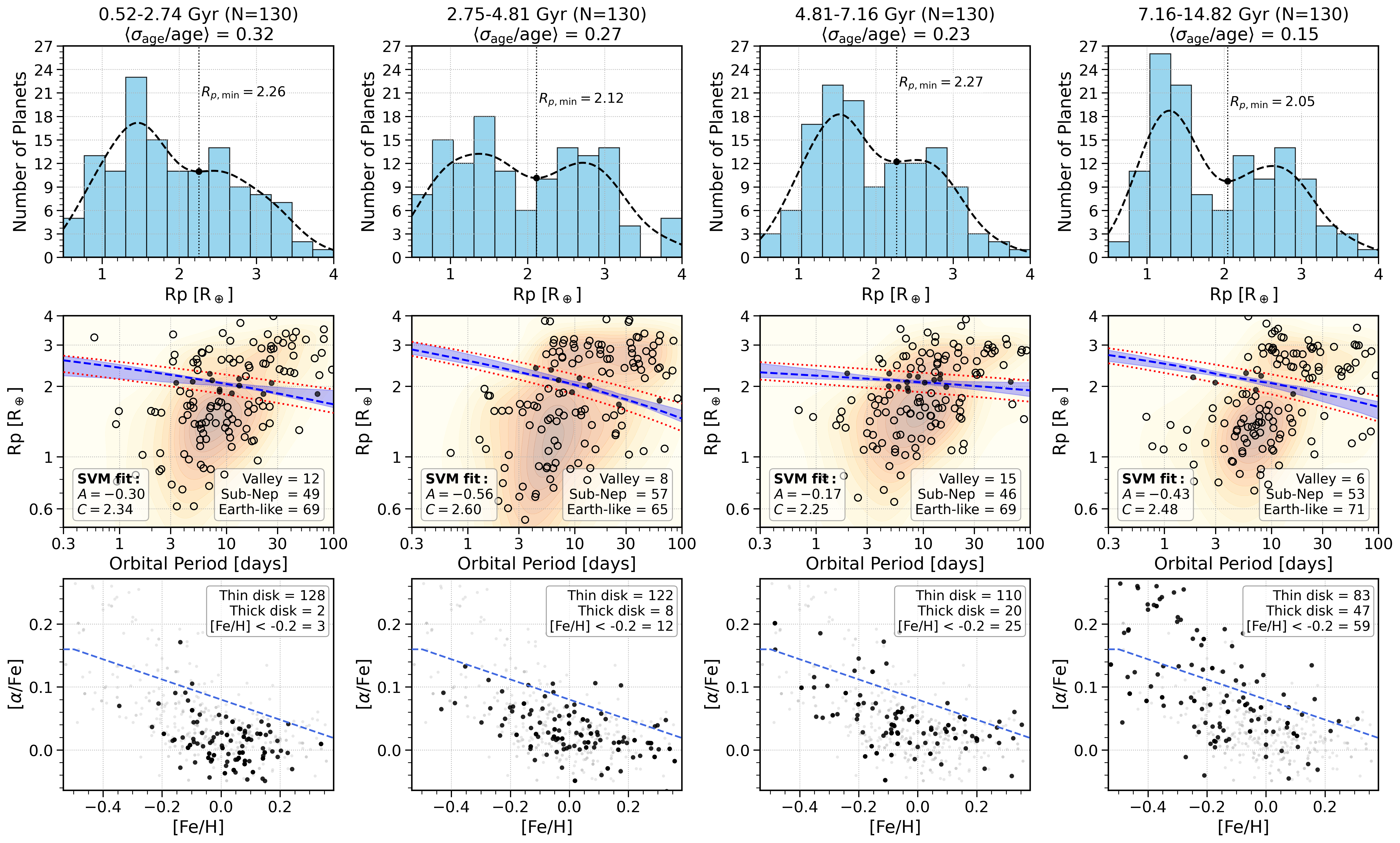}
\caption{
Comparison of planet radius distributions and radius--period diagrams across stellar age bins.
\textbf{Upper panels:} Planet-radius histograms for four equally populated age bins, defined by the 25th, 50th, and 75th percentiles of the stellar age distribution. The dashed curves show the corresponding kernel density estimates, and the inferred radius-valley locations from the KDE fits are indicated.
\textbf{Middle panels:} Radius--period diagrams for the same age bins, where the background shading represents the two-dimensional kernel density estimate in the $R_p$--$P_{\rm orb}$ plane. The dashed blue line indicates the median SVM-derived radius valley, the blue band shows the $1\sigma$ uncertainty from bootstrap resampling, and the dotted red curves mark a fixed $\pm0.2\,R_\oplus$ interval used to define planets within the valley (filled symbols); planets above and below the valley are shown as open symbols.
\textbf{Bottom panels:} Stellar chemical abundance distributions in the [$\alpha$/Fe]--[Fe/H] plane for the same age bins. Gray points show the full background sample, while black points indicate stars hosting planets in each age bin. The dashed line marks the adopted chemical separation between thin- and thick-disk populations: [$\alpha$/Fe]$=0.16$ for [Fe/H]$\leq -0.5$, and [$\alpha$/Fe]$=-0.16\times[\mathrm{Fe/H}]+0.08$ for [Fe/H]$>-0.5$. Stars above the boundary are classified as thick-disk and those below as thin-disk. The numbers of thin-disk and thick-disk stars, as well as the number of stars with [Fe/H]$<-0.2$, are indicated in each panel.
}
\label{fig:withage}
\end{figure*}

\begin{figure}
    \centering
    \includegraphics[width=\linewidth]{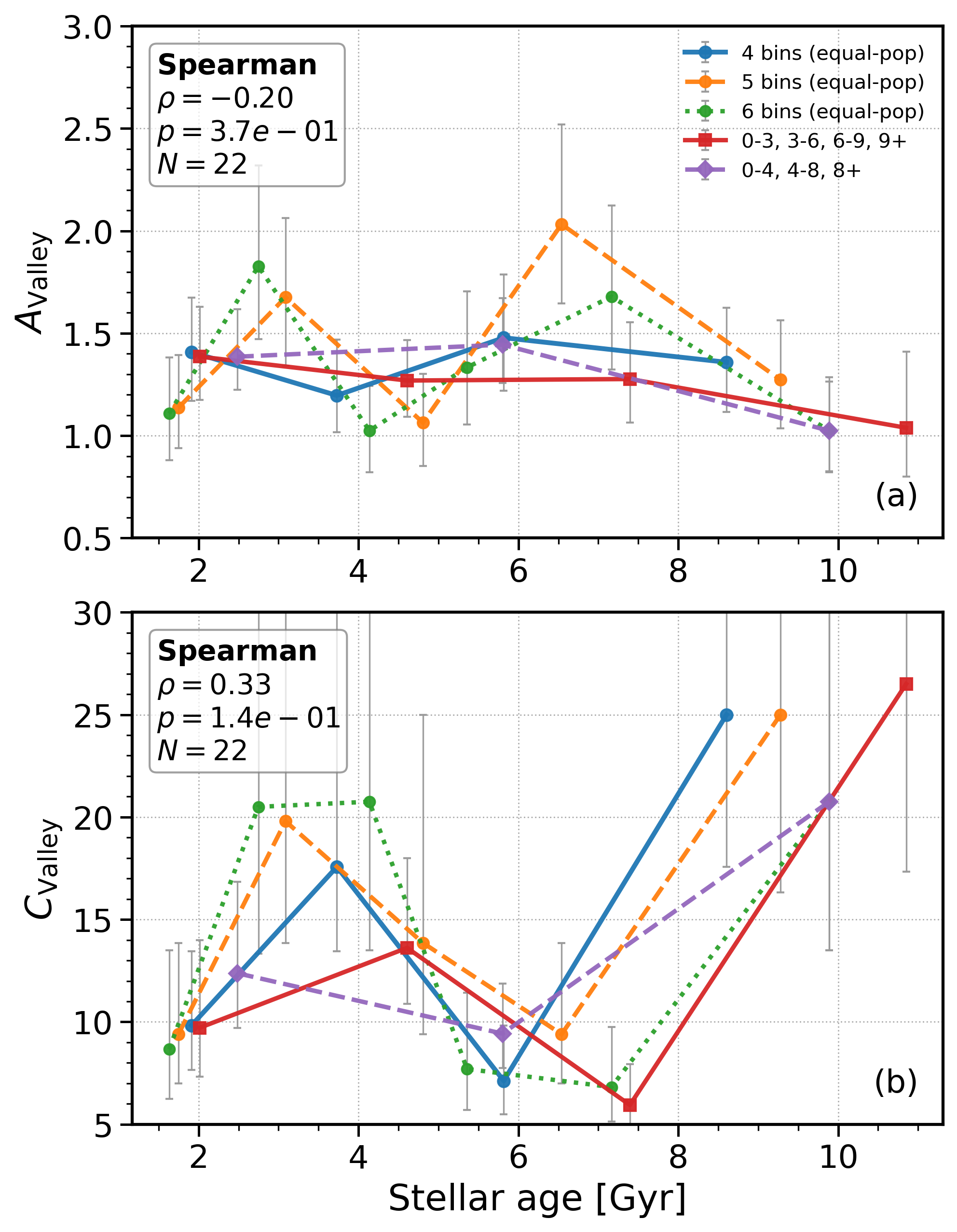}
    \caption{Evolution of planet-number ratios with stellar age. Panel~(a) shows the ratio of Earth-like planets to Sub-Neptune planets ($A_{\rm valley}$), while panel~(b) presents the ratio of non-valley planets (Earth-like + Sub-Neptune) to valley planets ($C_{\rm valley}$). Three curves correspond to equal-population age bins (4, 5, and 6 bins), while two curves show results from fixed-interval age bins. Error bars indicate the 16th--84th percentile range from bootstrap resampling. The Spearman correlation coefficients shown in the panels are computed by pooling the results from the different binning schemes and are intended to quantify the robustness of the trends rather than to represent independent statistical tests.}  
    \label{fig:parawithage}
\end{figure}

Stellar age provides a direct temporal probe of the physical origin of the planetary radius valley. Leading atmospheric-loss models predict different evolutionary timescales: photoevaporation is expected to operate primarily at early times under strong high-energy irradiation \citep[e.g.,][]{2013ApJ...775..105O,2014ApJ...792....1L,2017ApJ...847...29O,2021MNRAS.508.5886R}, whereas core-powered mass loss can continue over gigayear timescales as planets cool and contract \citep[e.g.,][]{2016ApJ...825...29G,2020MNRAS.493..792G}. If atmospheric erosion plays a dominant role, the prominence and morphology of the radius valley should therefore evolve with stellar age. Investigating the age dependence of the radius valley thus offers a direct observational test of the timescale and mechanism of planetary atmospheric loss.

It is important to emphasize that stellar age is not an independent control parameter, but a cumulative tracer of stellar and Galactic evolution, encoding correlated variations in chemical abundances, kinematics, and magnetic activity. As a result, any age dependence of the radius valley inevitably reflects the combined influence of multiple stellar properties. Although one could in principle attempt to control for
individual parameters such as [Fe/H] or [$\alpha$/Fe] \citep[e.g.,][]{2021AJ....162..100C}, doing so would substantially reduce the sample size within each age bin and compromise the robust identification of the radius valley, which is intrinsically a low-occupancy feature. We therefore focus on characterizing how the valley morphology varies with age in the full sample, and subsequently examine its behavior across other stellar parameters in separate analyses. This approach is further motivated by the fact that even within the Galactic thin disk, stellar populations do not follow a single, monotonic evolutionary pathway: while the classical picture distinguishes an old, $\alpha$-enhanced thick disk formed on a short timescale from a younger, low-$\alpha$ thin disk formed more quiescently, recent studies increasingly suggest that the thin disk itself has experienced a complex evolution, potentially shaped by minor merger--induced star formation episodes within the past $\sim$4~Gyr \citep{2020NatAs...4..965R,2025NatCo..16.1581S}. Consistent with this view, the age--metallicity distribution of the total sample exhibits a pronounced ``V-shape'' \citep{2018MNRAS.477.2326F,2019MNRAS.489.1742F,2022MNRAS.512.2890L,2022Natur.603..599X,2023MNRAS.523.1199S,2025ApJ...995...33C} for stars younger than $\sim$6~Gyr. This chemically and dynamically complex structure of the thin disk reinforces the view that stellar age—along with the parameters examined in the following sections—serves as an integrated evolutionary coordinate, rather than a quantity that can be cleanly isolated from other stellar properties in the primary analysis. To compare the radius-valley morphology along each parameter while maintaining comparable sample sizes and sufficient statistical robustness, we divide the sample into equally populated bins defined by the 25th, 50th, and 75th percentiles of the parameter distribution. This procedure orders the subsamples from lower to higher values of the parameter and allows us to examine the monotonic variation of the valley morphology within the sample. The same binning scheme is applied consistently to all stellar parameters examined in this work.

Figure~\ref{fig:withage} illustrates the evolution of the planet radius distribution as a function of stellar age. The sample is sorted by stellar age and divided into four equal-population bins defined by the 25th, 50th, and 75th percentiles. The binning uses the nominal ages only (age uncertainties are not included). The average relative age uncertainty in each bin is shown in the panel titles. The top row presents the one-dimensional planet-radius histograms, the middle row shows the radius--period ($R_p$--$P_{\rm orb}$) distributions with kernel density contours and the SVM-derived valley locations, and the bottom row displays the corresponding stellar chemical distributions in the [$\alpha$/Fe]--[Fe/H] plane. The top row presents the one-dimensional planet radius histograms, the middle row shows the radius--period ($R_p$--$P_{\rm orb}$) distributions with kernel density contours and SVM-derived valley locations, and the bottom row displays the corresponding stellar chemical distributions in the [$\alpha$/Fe]--[Fe/H] plane. In the youngest subsample (0.52--2.74~Gyr), the radius valley is comparatively weak. The radius histogram does not show a clear deficit, and the two-dimensional density contours in the $R_p$--$P_{\rm orb}$ plane lack a well-defined bimodality. In the subsequent 2.75--4.81~Gyr bin, a distinct valley emerges near $\sim$2~$R_\oplus$: the density contours reveal a clear separation between two planet populations, accompanied by a strong period dependence of the valley location (slope $\simeq -0.56$), and the numbers of Earth-like and sub-Neptune planets become comparable. Interestingly, the valley signature weakens again in the intermediate-age bin (4.81--7.16~Gyr). In this subsample, the deficit in the one-dimensional radius distribution becomes less pronounced, the SVM-defined valley region is substantially refilled (with the number of planets inside the valley increasing from 8 to 15), and the dependence of the valley location on orbital period is significantly reduced (slope $\simeq -0.17$). In contrast, the oldest subsample (7.16--14.82~Gyr) shows a renewed strengthening of the valley. A pronounced deficit is visible at $R_p \simeq 1.5$--2~$R_\oplus$ in the radius histogram, Earth-like planets become significantly more abundant than sub-Neptunes, the SVM-defined valley region is again sparsely populated (with the number of planets decreasing from 15 to 6), and the valley location recovers a strong period dependence (slope $\simeq -0.43$). The increasing fraction of Earth-like planets with stellar age is broadly consistent with previous studies suggesting that metal-poor environments preferentially form or preserve smaller planets \citep{2012Natur.486..375B,2012A&A...545A..32A,2025ApJ...995...33C}. As older stellar populations are typically more metal-poor, this demographic shift naturally enhances the contrast of the radius valley, making it most prominent in the oldest age bin. 

In our sample, the youngest age bin does not exhibit a clear radius valley. A natural interpretation is that a significant fraction of planetary systems in this age range have not yet experienced sufficient atmospheric mass loss or internal cooling to develop a clear valley structure. The subsequent delayed emergence, intermediate weakening, and later re-strengthening of the valley contrast over gigayear timescales are more naturally explained by models in which atmospheric loss proceeds gradually as planets cool and contract, as in core-powered mass-loss scenarios \citep[e.g.,][]{2013ApJ...775..105O,2014ApJ...795...65J}. By contrast, photoevaporation models generally predict a rapid establishment of the radius valley within the first few hundred Myr to $\sim$1~Gyr \citep[e.g.,][]{2016ApJ...825...29G,2019MNRAS.487...24G,2020MNRAS.493..792G}, making it more challenging to reconcile them with the absence of a well-defined valley in our youngest subsample. This behavior differs from the result of \citet{David_2021}, who reported a strongly depleted valley at ages younger than $\sim$2--3~Gyr. However, their age bins are not directly comparable to ours, as the $\sim$2--3~Gyr interval spans two of our bins (0.52--2.74 and 2.75--4.81~Gyr). The difference may therefore reflect the different binning schemes, the large age uncertainties at young ages (typical relative errors of $\sim$32\%), and differences in sample selection.

The stellar chemical distributions shown in the bottom row provide a natural explanation for the non-monotonic evolution of the radius valley with stellar age. Following the chemical classification of \citet{2022Natur.603..599X}, we separate thin- and thick-disk stars using their positions in the [$\alpha$/Fe]--[Fe/H] plane. In the youngest bin (0.52--2.74~Gyr), the radius valley is weak, likely because atmospheric loss processes have not yet fully sculpted the planet radius distribution. In the 2.75--4.81~Gyr bin, the stellar population is overwhelmingly dominated by chemically homogeneous thin-disk stars ($\sim$94\% thin disk), and the radius valley becomes clearly visible. In the 4.81--7.16~Gyr bin, the thin-disk fraction decreases to $\sim$84\%, placing this subsample near the transition region between the thin and thick disk. The resulting mixture of stellar populations weakens the coherence of the planet radius distribution and makes the radius valley less distinct. In the oldest bin (7.16--14.82~Gyr), the contribution from thick-disk stars increases substantially (83 thin-disk and 47 thick-disk stars), while the fraction of metal-poor stars ([Fe/H]$<-0.2$) rises from 19\% to 45\%. In this regime the radius valley becomes most prominent. Taken together, these results suggest that the observed morphology of the radius valley is strongly influenced by stellar population mixing. The apparent weakening of the valley at intermediate ages does not reflect a reversal in planetary evolutionary processes, but instead arises from the superposition of Galactic stellar populations with different chemical and dynamical histories.

Figure~\ref{fig:parawithage} quantifies the evolution of the radius-valley morphology as a function of stellar age through the parameters $A_{\rm valley}$ and $C_{\rm valley}$. To assess the robustness of the inferred trends, we repeat the analysis using equal-population binning schemes with 4, 5, and 6 age bins, and additionally test several fixed age-bin definitions to examine the sensitivity of the results to the binning strategy. All schemes yield consistent evolutionary patterns, demonstrating that the observed behavior is not driven by the specific choice of binning. Panel~(a) shows that $A_{\rm valley}$ exhibits a clearly non-monotonic evolution with age. It is lowest at the youngest ages, increases toward $\sim$3~Gyr, declines around $\sim$4--5~Gyr, rises again to a secondary maximum near $\sim$6--7~Gyr, and subsequently decreases toward older ages. This behavior likely reflects the combined influence of planetary evolution and stellar population mixing. The initial rise is consistent with the gradual emergence of smaller planets as atmospheric mass-loss processes operate over gigayear timescales, whereas the decline at intermediate ages coincides with the increasing contribution of mixed thin- and thick-disk stellar populations, which dilutes the contrast between distinct planet-size populations. The secondary enhancement at $\sim$6--7~Gyr may be associated with the growing contribution of older, typically more metal-poor stars, consistent with previous findings that the super-Earth to sub-Neptune ratio increases over gigayear timescales \citep{2021ApJ...911..117S}. Panel~(b) shows a closely related evolution in $C_{\rm valley}$. The valley contrast is weakest at young ages ($\lesssim$2~Gyr), strengthens toward $\sim$4~Gyr, weakens again around $\sim$6--8~Gyr, and increases steadily at older ages. This trend mirrors the morphological changes seen in Figure~\ref{fig:withage} and reinforces the conclusion that the radius valley is least pronounced both in newly formed planetary systems and in age intervals dominated by chemically mixed stellar populations, while it becomes most prominent in relatively homogeneous stellar populations or in systems formed in more metal-poor environments.

Taken together, the evolution of the radius valley with stellar age is clearly non-monotonic and reflects more than a simple temporal sequence. Its prominence and morphology are shaped by the interplay between planetary radius evolution and the changing mixture of Galactic stellar populations. At young ages, the absence of a well-defined valley is consistent with incomplete atmospheric loss or core cooling, while at intermediate ages the superposition of chemically and dynamically distinct thin- and thick-disk populations weakens the valley signature. Only at older ages, where the sample is dominated by a relatively homogeneous, metal-poor population, does the radius valley reach its maximum contrast. These results indicate that stellar age alone is an insufficient descriptor of the processes shaping the radius valley; instead, age acts as an integrated proxy for stellar chemistry and formation environment. In this sense, the radius valley encodes not only planetary atmospheric evolution, but also the chemo-dynamical assembly history of the Galactic disk.

\subsection{The Radius Valley across Stellar Magnetic Activity}

\begin{figure*}
    \centering
    \includegraphics[width=\linewidth]{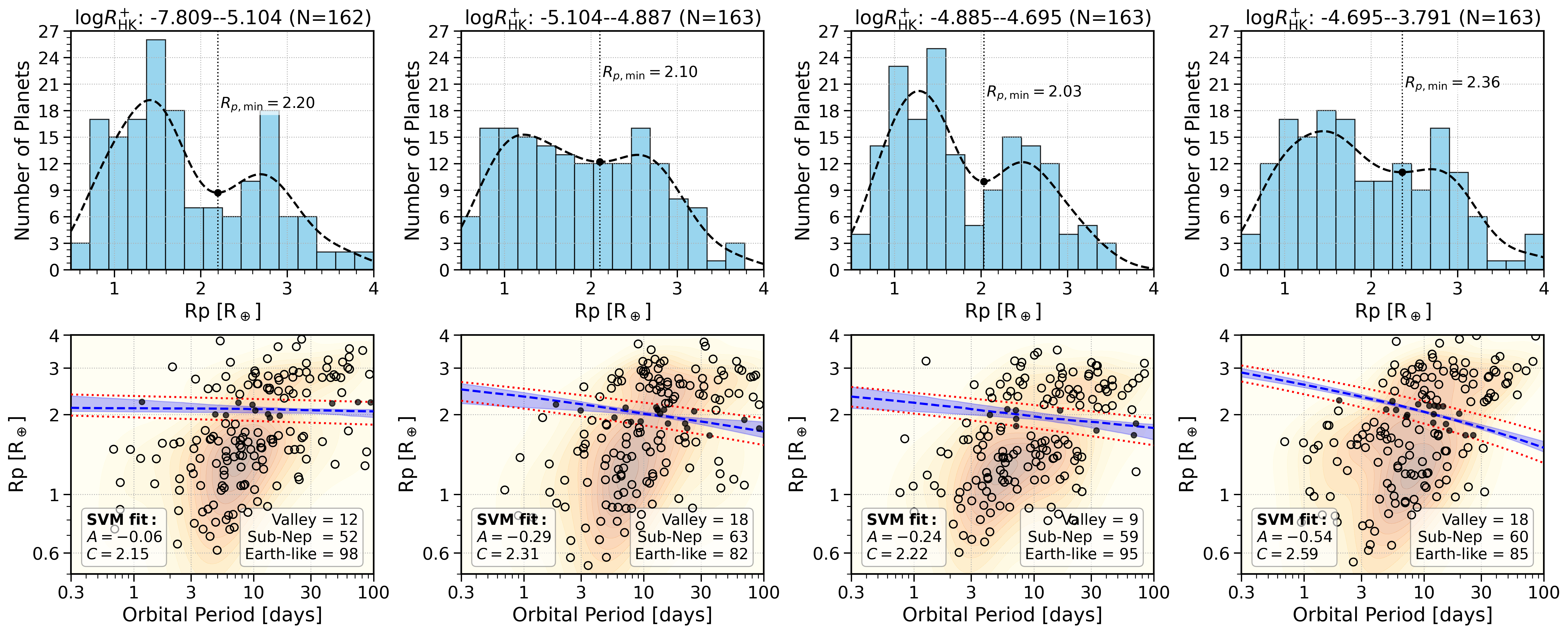}
    \caption{
    Same as Figure~\ref{fig:withage}, but binned by stellar magnetic activity
    quantified using $\log R^+_{\rm HK}$.}
    \label{fig:withlogRHK}
\end{figure*}

\begin{figure}
    \centering
    \includegraphics[width=\linewidth]{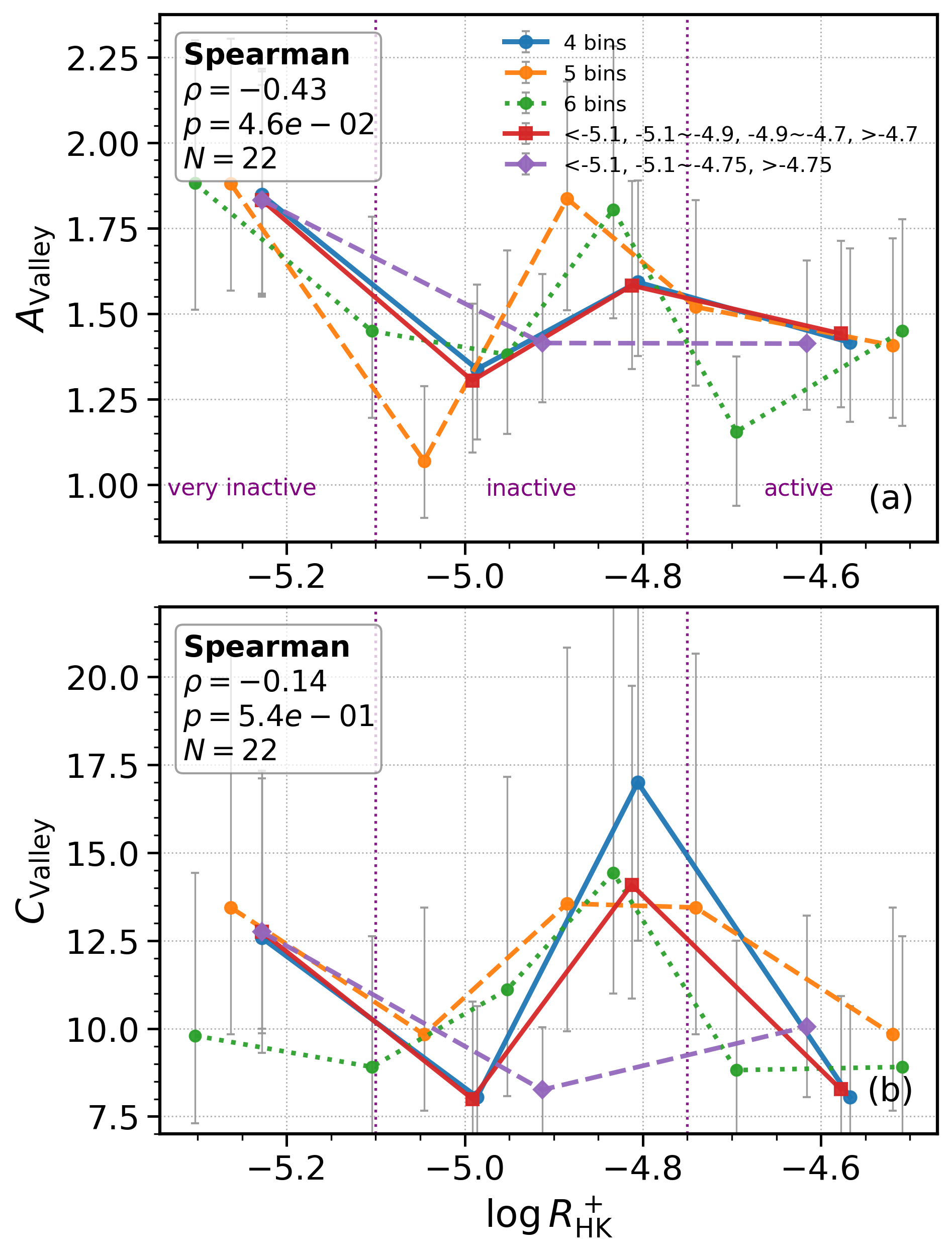}
    \caption{Evolution of planet-number ratios with stellar magnetic activity, quantified by $\log R^+_{\rm HK}$. Symbols are similar as those in Figure~\ref{fig:parawithage}. The purple line shows the results obtained using fixed activity intervals motivated by the chromospheric activity regimes defined in \citet{1996AJ....111..439H}, corresponding to the very inactive, inactive, and active regimes in our sample.}
    \label{fig:parawithlogRHKplus}
\end{figure}

Stellar magnetic activity is expected to influence the evolution of close-in planets by tracing the high-energy radiative and magnetically driven environment experienced by planetary atmospheres, and has therefore been discussed in the context of planet formation and evolution \citep{2012MNRAS.425.2931O,2021AJ....162..100C}. In particular, high-energy (X-ray and extreme-ultraviolet) emission associated with magnetic activity provides the primary energy source for photoevaporative mass loss, whereas core-powered mass-loss scenarios are largely insensitive to the stellar high-energy environment. The dependence of the radius valley on stellar activity therefore offers a potential diagnostic for distinguishing between these mechanisms. In this work, we examine the radius valley as a function of stellar chromospheric activity using the photosphere-corrected indicator $\log R^+_{\rm HK}$, which provides a bolometric-normalized measure of Ca~{\sc ii} H\&K emission and enables a physically uniform comparison across different stellar types.

Figure~\ref{fig:withlogRHK} examines the morphology of the radius valley as a function of $\log R^+_{\rm HK}$. The sample is sorted by activity level and divided into four equal-population bins spanning magnetically quiet to highly active stars. In the lowest-activity bin, the planet population is dominated by Earth-like planets, yielding the highest Earth-like fraction ($\sim$60\%) among all activity groups. A clear deficit is visible at $R_p \simeq 1.8$--2.2~$R_\oplus$, indicating a well-defined radius valley, while the SVM fit shows only a weak dependence of the valley location on orbital period (slope $\simeq -0.06$). As $\log R^+_{\rm HK}$ increases to the second bin, the valley becomes much less pronounced. The deficit near $\sim$2~$R_\oplus$ largely disappears, and the SVM-defined valley region is substantially refilled, containing 18 planets, indicating a weak or poorly defined bimodality. With a further increase in activity to the third bin, the radius valley re-emerges most clearly. The deficit around $\sim$2~$R_\oplus$ becomes prominent again, the SVM-defined valley region reaches its lowest occupancy (9 planets), and the period dependence of the valley location strengthens (slope $\simeq -0.24$). Interestingly, at the highest activity levels (fourth bin), the valley becomes less distinct once more. No clear deficit is observed near $\sim$2~$R_\oplus$, and the SVM-defined valley region is again substantially refilled (18 planets). Despite this, the correlation between valley location and orbital period is strongest in this bin.

Figure~\ref{fig:parawithlogRHKplus} further quantifies the activity dependence of the radius valley using the number ratios $A_{\rm valley}$ and $C_{\rm valley}$ as functions of $\log R^+_{\rm HK}$. In addition to the equal-population bins, we also examine fixed activity intervals motivated by the chromospheric activity regimes defined by \citet{1996AJ....111..439H}. Because our activity indicator is $\log R^+_{\rm HK}$ rather than the classical $\log R'_{\rm HK}$, we adopt the same numerical boundaries only as an approximate reference: very inactive ($\log R^+_{\rm HK}<-5.1$), inactive ($-5.1<\log R^+_{\rm HK}<-4.75$), active ($-4.75<\log R^+_{\rm HK}<-4.2$), and very active ($\log R^+_{\rm HK}>-4.2$). Applying this scheme yields 165 very inactive stars, 287 inactive stars, 190 active stars, and only 9 very active stars, indicating that most planet-hosting stars fall in the inactive regime, consistent with \citet{2025ApJS..281...61L}. Because the very active subsample is extremely small, the fixed-bin results (purple curve in Figure~\ref{fig:parawithlogRHKplus}) effectively trace the very inactive, inactive, and active regimes. Within this classification, both $A_{\rm valley}$ and $C_{\rm valley}$ reach their highest values in the very inactive regime, indicating that the contrast between Earth-like and Sub-Neptune planets is strongest among the least active stars. Toward higher activity levels (inactive and active regimes), both ratios generally decrease, implying that the radius valley becomes progressively less pronounced as stellar activity increases. The equal-population binning analysis shows a qualitatively similar overall behavior: both $A_{\rm valley}$ and $C_{\rm valley}$ are relatively high at the lowest activity levels, show a secondary enhancement around $\log R^+_{\rm HK}\sim -4.9$ to $-4.8$, and reach their lowest values at the highest activity levels.

These results demonstrate that stellar magnetic activity does not modulate the radius valley in a simple or monotonic manner. In photoevaporation-driven scenarios, stronger magnetic activity—serving as a proxy for enhanced high-energy (X-ray and extreme-ultraviolet) irradiation—is expected to increase atmospheric mass-loss efficiency and thereby deepen the depletion near the valley \citep[e.g.,][]{2013ApJ...775..105O,2021MNRAS.508.5886R}. However, we do not observe a systematic increase in valley contrast with increasing $\log R^+_{\rm HK}$. Instead, both $A_{\rm valley}$ and $C_{\rm valley}$ show a non-monotonic dependence on activity. In particular, when adopting the chromospheric activity regimes defined by \citet{1996AJ....111..439H}, the radius valley is most prominent in the very inactive regime ($\log R^+_{\rm HK}<-5.1$), while the contrast weakens toward the active regimes. The weak and non-monotonic dependence of valley depth on stellar activity suggests that high-energy irradiation is not the primary factor controlling the overall depletion between the two planet populations. This behavior is more naturally consistent with core-powered mass-loss scenarios, in which atmospheric escape is driven predominantly by the thermal evolution of the planet and is largely insensitive to the stellar high-energy environment. Interestingly, the SVM-derived valley slope becomes steeper toward higher activity levels. This behavior suggests that stellar magnetic activity primarily reshapes the geometry of the radius valley—strengthening its period dependence—rather than enhancing its global contrast.

\subsection{The Radius Valley across Stellar Elemental Abundances}
Stellar elemental abundances set fundamental conditions for planet formation and subsequent evolution by regulating both the solid content of protoplanetary disks and the internal composition of planets. It is well established that the occurrence rate of giant planets increases with host-star metallicity \citep[e.g.,][]{2004A&A...415.1153S,2005ApJ...622.1102F,2010PASP..122..905J}, with the dependence being strongest for Jovian-mass planets \citep[e.g.,][]{2005ApJ...622.1102F}. Beyond overall metallicity, abundance ratios such as C/O and Mg/Si influence the availability of condensable material and the mineralogy of rocky cores \citep{2016ApJ...831...20B}, thereby affecting core masses, bulk densities, and the ability of planets to accrete and retain gaseous envelopes. Stellar chemistry is therefore expected to influence not only planet formation, but also the processes shaping the radius valley. In core-powered mass-loss scenarios, the efficiency of atmospheric loss depends sensitively on core mass and internal structure, both of which are linked to the solid inventory set by the disk composition. By contrast, photoevaporation is primarily controlled by the high-energy stellar irradiation and is only indirectly affected by stellar abundances through their influence on planetary gravity. The dependence of the radius-valley morphology on stellar chemical properties thus provides an important diagnostic for distinguishing between formation-controlled and irradiation-driven origins of the valley.

\begin{figure*}
    \centering
    \includegraphics[width=\linewidth]{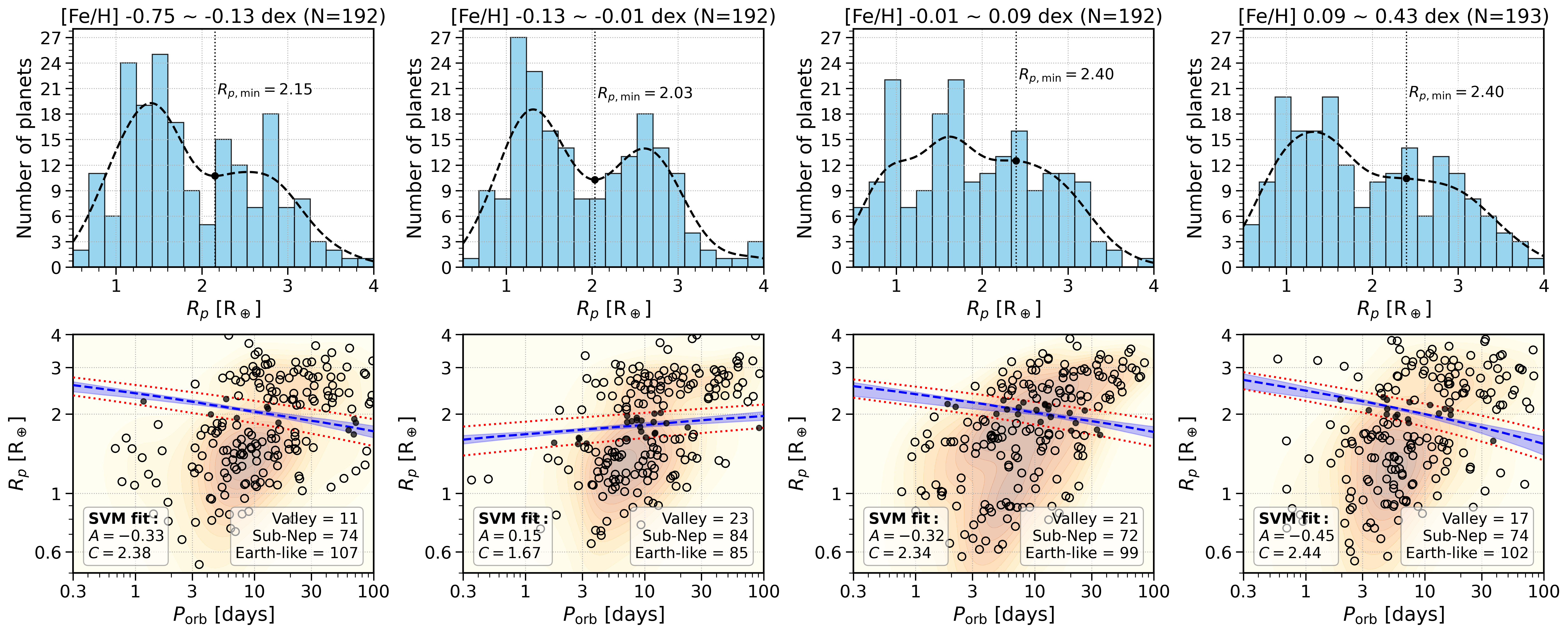}
    \caption{
    Same as Figure~\ref{fig:withage}, but binned by stellar metallicity ([Fe/H]).}
    \label{fig:withfeh}
\end{figure*}

\begin{figure}
    \centering
    \includegraphics[width=\linewidth]{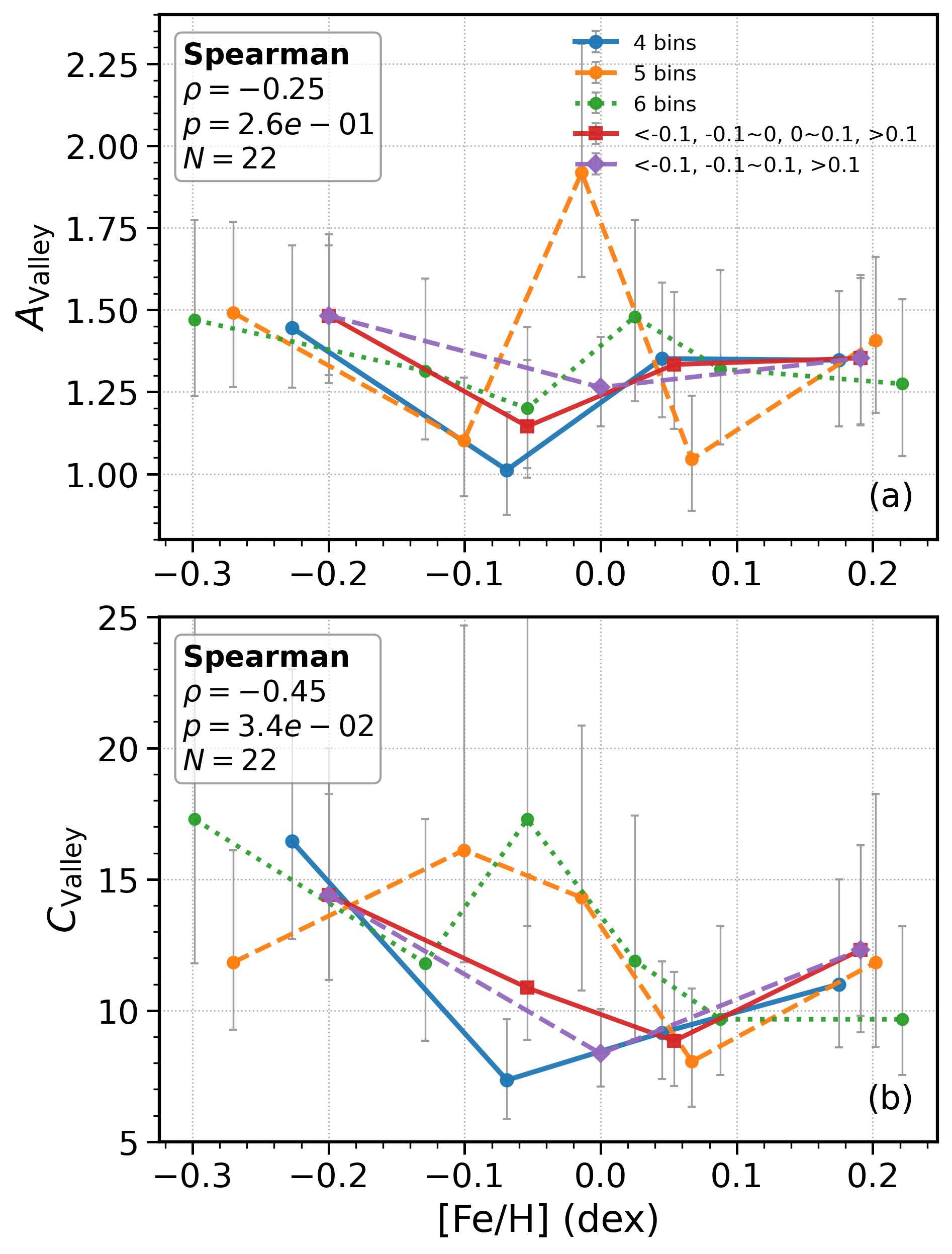}
    \caption{
    Evolution of planet-number ratios with stellar metallicity ([Fe/H]). Symbols are similar as those in Figure \ref{fig:parawithage}.}
    \label{fig:parawithfeh}
\end{figure}

Figure~\ref{fig:withfeh} shows that the morphology of the radius valley varies systematically with stellar metallicity. In the most metal-poor bin, a well-defined valley is present near $R_p \sim 2.2\,R_\oplus$, characterized by the largest number of Earth-like planets (107 planets), the lowest occupancy within the SVM-defined valley region (11 planets), and a strong period dependence of the valley location (slope $\simeq -0.33$), indicating a clearly separated bimodal distribution. At intermediate metallicity ($-0.13 \lesssim {\rm [Fe/H]} \lesssim -0.01$), a deficit near $\sim$2~$R_\oplus$ remains visible, but the valley becomes less distinct. The SVM-defined region is substantially refilled (23 planets), and the period dependence weakens and even reverses (slope $\simeq 0.15$). In the near-solar metallicity bin ($-0.01 \lesssim {\rm [Fe/H]} \lesssim 0.09$), the bimodality is further reduced, with only a weak deficit near $\sim$2~$R_\oplus$ and significant filling of the valley region (21 planets). In the most metal-rich bin, a deficit is still present near $R_p \sim 2,R_\oplus$, but an additional depletion emerges around $R_p \sim 2.6,R_\oplus$, leading to a broader and less coherent bimodal structure. The SVM-defined valley region remains relatively populated (17 planets), implying a reduced contrast between the two planet populations.

Figure~\ref{fig:parawithfeh} further quantifies these trends using the ratios $A_{\rm valley}$ and $C_{\rm valley}$. Both parameters exhibit a non-monotonic dependence on stellar metallicity. $A_{\rm valley}$ is relatively high at the lowest metallicities, decreases toward ${\rm [Fe/H]}\sim-0.1$, and then rises again with increasing metallicity. Likewise, $C_{\rm valley}$ reaches its highest value in the most metal-poor regime, declines toward ${\rm [Fe/H]}\sim0.1$, and then increases again at higher metallicities.

Stellar metallicity is generally regarded as a reliable tracer of the primordial disk metallicity \citep{2003A&A...398..363S}. The observed metallicity dependence therefore likely reflects changes in the underlying planet population rather than a direct shift of the valley locus itself. In particular, population-synthesis models demonstrate that disk metallicity imprints on overall planet demographics \citep{2012A&A...541A..97M}. Furthermore, theoretical studies have shown that the contrast between the two radius peaks—hence the apparent depth of the valley—is sensitive to the underlying planet-mass distribution, even when the valley location remains largely unchanged \citep{2019MNRAS.487...24G}. In this context, the relatively strong valley at low metallicities suggests that metal-poor systems preferentially host planets with smaller core masses, leading to a more strongly depleted valley region. Toward near-solar metallicities, the broader range of core masses expected in metal-rich disks can produce a wider dispersion in envelope retention efficiencies, partially filling the valley and reducing its contrast. At the highest metallicities, the valley contrast appears to increase again, possibly reflecting a shift toward more massive cores and more efficient atmospheric loss for a subset of planets. Such a pronounced metallicity dependence is not naturally expected in photoevaporation scenarios, where atmospheric loss is primarily regulated by stellar high-energy irradiation and depends only weakly on stellar composition. In contrast, core-powered mass-loss models predict a strong sensitivity to core mass and internal structure, both of which are set by the disk solid content and therefore correlate with stellar metallicity. The observed trend thus points to a dominant role for core-controlled processes in shaping the radius valley.

\begin{figure*}
    \centering
    \includegraphics[width=\linewidth]{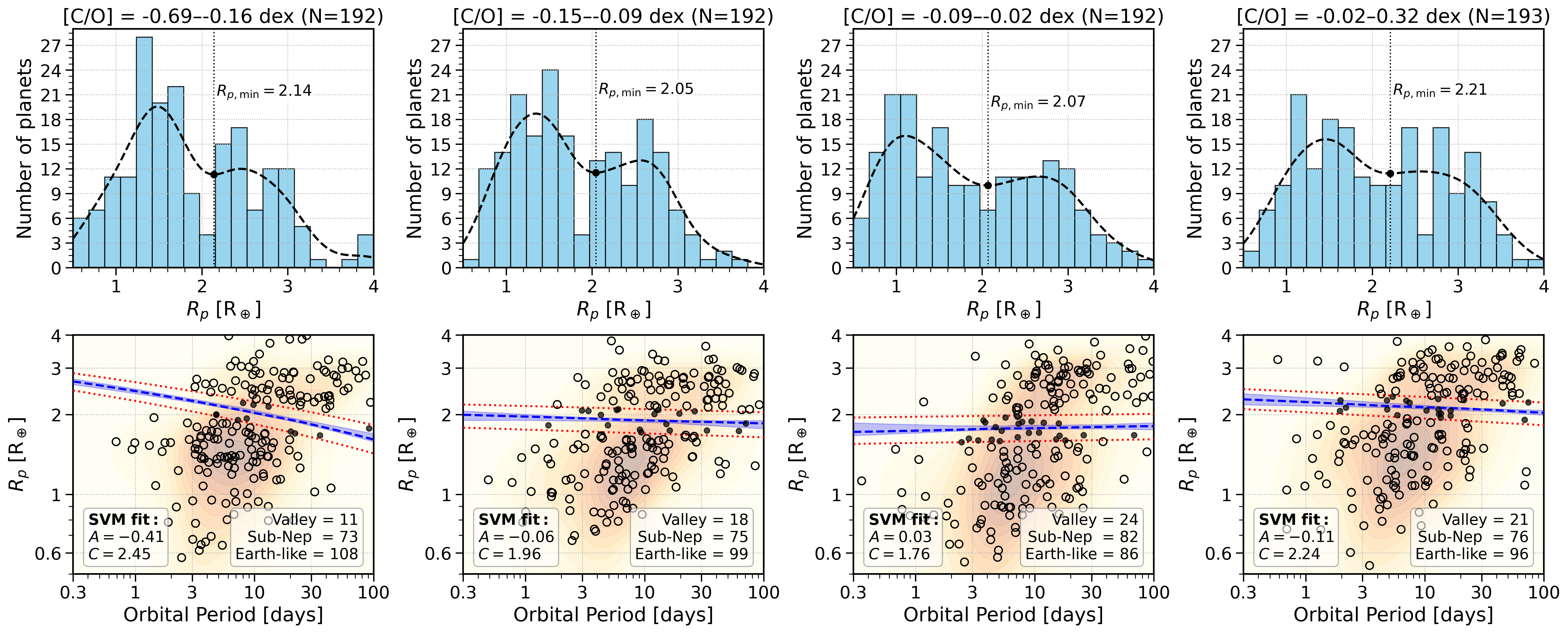}
    \caption{
    Same as Figure~\ref{fig:withage}, but binned by stellar [C/O].}
    \label{fig:withCO}
\end{figure*}

\begin{figure}
    \centering
    \includegraphics[width=\linewidth]{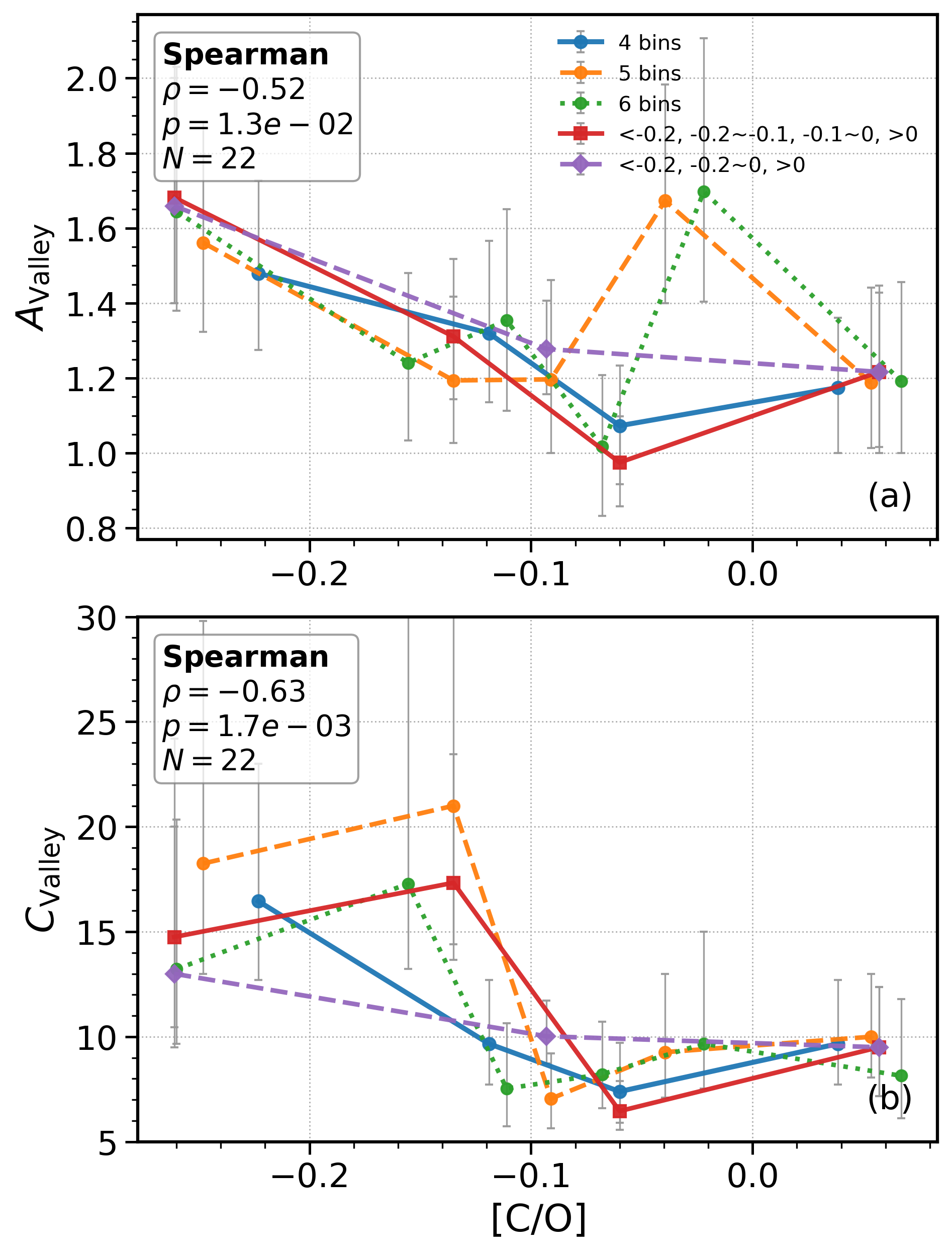}
    \caption{
    Evolution of planet-number ratios with stellar [C/O]). Symbols are similar as those in Figure \ref{fig:parawithage}.}
    \label{fig:parawithCO}
\end{figure}

\begin{figure*}
    \centering
    \includegraphics[width=\linewidth]{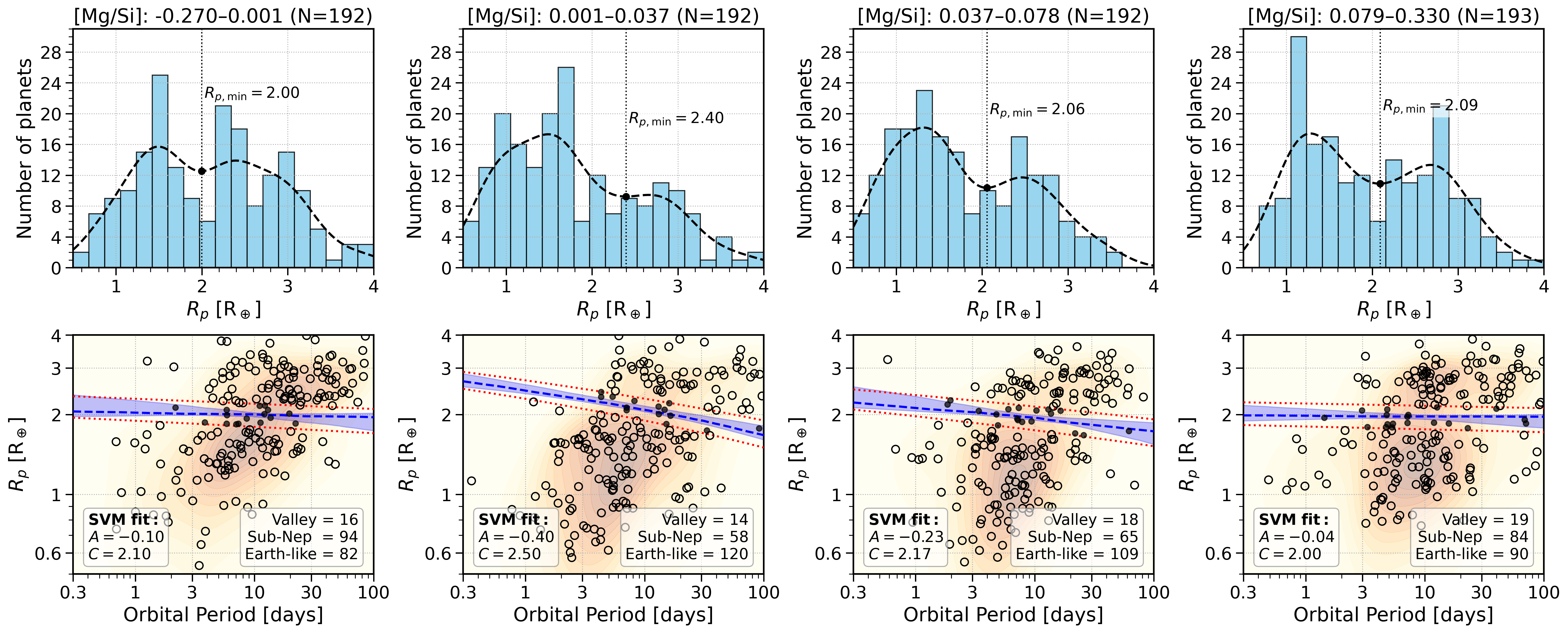}
    \caption{
    Same as Figure~\ref{fig:withage}, but binned by stellar [Mg/Si].}
    \label{fig:withMgSi}
\end{figure*}

\begin{figure}
    \centering
    \includegraphics[width=\linewidth]{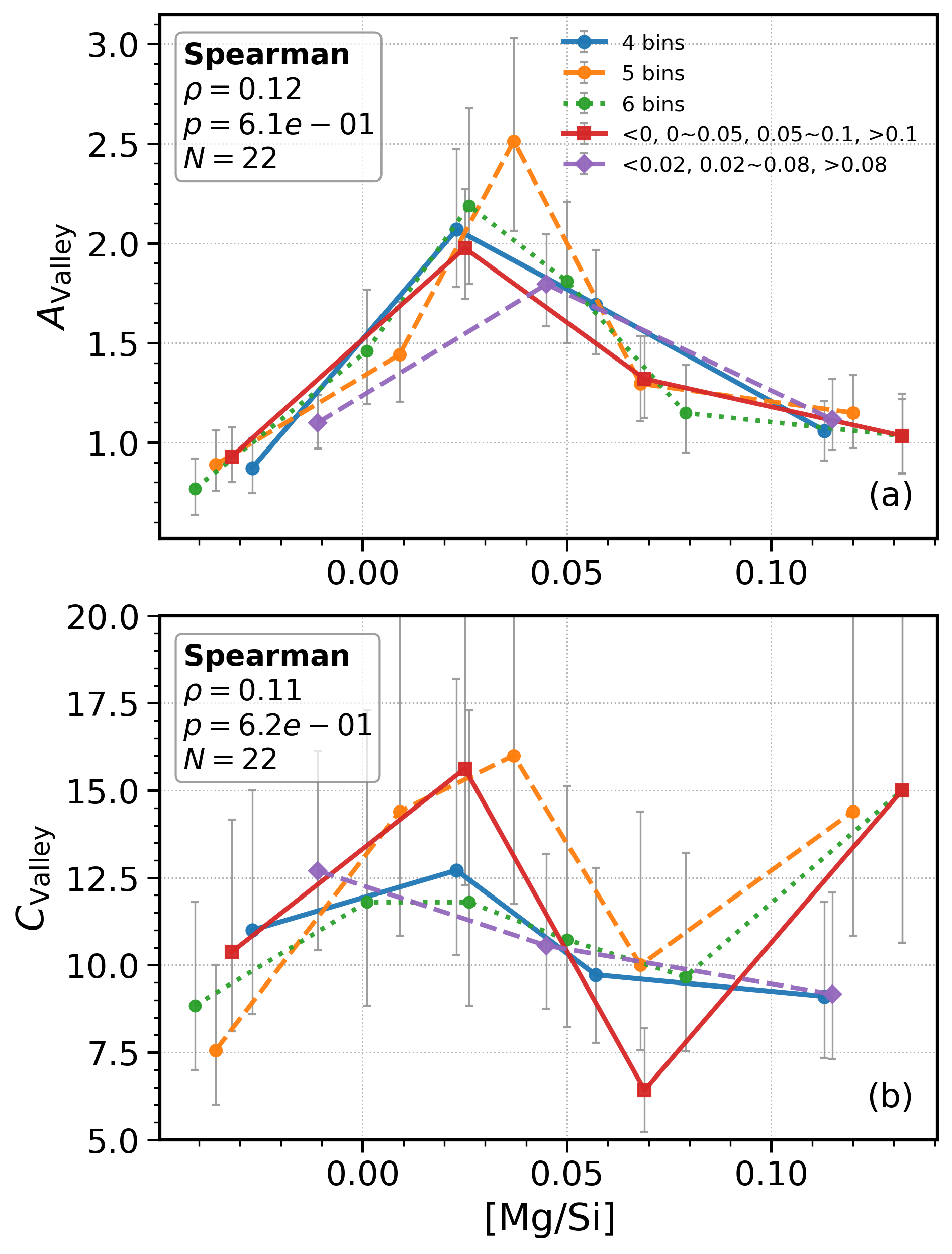}
    \caption{
    Evolution of planet-number ratios with stellar [Mg/Si]). Symbols are similar as those in Figure \ref{fig:parawithage}.}
    \label{fig:parawithMgSi}
\end{figure}

Figures~\ref{fig:withCO} shows that the morphology of the radius valley exhibits a dependence on stellar [C/O]. The lowest-[C/O] bin displays the most pronounced radius valley, characterized by a strong deficit of planets near $R_p \simeq 2.1\,R_\oplus$ and by the smallest number of planets (11 planets) residing within the valley region. As [C/O] increases, the region around $\sim2\,R_\oplus$ becomes progressively more populated, and the radius valley correspondingly appears less distinct. This behavior is quantified in Figure~\ref{fig:parawithCO}. Although the detailed trends show some sensitivity to the adopted binning scheme, both the Earth-like fraction proxy $A_{\rm valley}$ and the valley-contrast proxy $C_{\rm valley}$ exhibit an overall decline with increasing [C/O]. These results indicate that the radius valley is clearest at low [C/O] and becomes progressively filled toward higher [C/O].

This trend is physically plausible in the context of disk chemistry and planetesimal composition. Carbon and oxygen dominate the molecular budget of protoplanetary disks, and the [C/O] ratio regulates the availability of free oxygen for silicate formation \citep{2014prpl.conf..363P, 2013pccd.book.....G}. For [C/O] ratios below the critical threshold, planetesimal geology is expected to be dominated by magnesium silicates, as in the Solar System \citep{2005astro.ph..4214K, 2014ApJ...787...81M, 2009ARA&A..47..481A}. Such compositions favor the formation of relatively low-mass rocky cores, which are more susceptible to atmospheric loss. Consequently, systems with low [C/O] naturally produce a sharper separation between bare rocky planets and envelope-bearing sub-Neptunes, leading to a more pronounced radius valley. As [C/O] increases, changes in planetesimal composition and core properties may promote partial atmospheric retention, progressively filling the valley region.

Figures~\ref{fig:withMgSi} shows how the morphology of the radius valley varies with stellar $\mathrm{[Mg/Si]}$. A deficit near $R_p \sim 2\,R_\oplus$ is present across all bins, but the relative population balance and valley structure exhibit systematic changes at higher $\mathrm{[Mg/Si]}$. In the three higher-$\mathrm{[Mg/Si]}$ bins ($\mathrm{[Mg/Si]} \gtrsim 0.001$), the fraction of Earth-like planets decreases while the number of sub-Neptunes increases (Earth-like: 120 → 109 → 90; sub-Neptune: 58 → 65 → 84). At the same time, the period dependence of the valley location weakens, with the SVM slope flattening from $-0.40$ to $-0.23$ and then to $-0.04$, indicating a progressively less well-defined valley structure. These trends are quantified in Figure~\ref{fig:parawithMgSi}. Both $A_{\rm valley}$ and $C_{\rm valley}$ increase from low $\mathrm{[Mg/Si]}$ to a peak at $\mathrm{[Mg/Si]}\simeq0.03$. At higher $\mathrm{[Mg/Si]}$, $A_{\rm valley}$ decreases monotonically, whereas $C_{\rm valley}$ declines to $\mathrm{[Mg/Si]}\sim0.08$ and shows a possible upturn at higher $\mathrm{[Mg/Si]}$.

The observed dependence of the radius-valley morphology on $\mathrm{[Mg/Si]}$ suggests that variations in magnesium–silicate planetesimal composition lead to systematic differences in rocky core structure and thermal evolution. Modest departures from the solar Mg/Si ratio alter the relative proportions of major silicate phases \citep{2012ApJ...747L...2C,2013pccd.book.....G}, potentially affecting core density, differentiation, and heat transport, and thereby regulating the efficiency of atmospheric cooling and escape. The sensitivity of the valley to $\mathrm{[Mg/Si]}$ therefore points to a dependence on planetary internal structure rather than stellar irradiation, and is more naturally consistent with core-powered mass-loss scenarios than with irradiation-driven photoevaporation. When combined with population mixing and other stellar properties that co-vary with $\mathrm{[Mg/Si]}$, such compositional effects can naturally account for the irregular and non-monotonic behavior observed.

Compared with [C/O], the trends associated with $\mathrm{[Mg/Si]}$ are more robust across different binning schemes, and the radius valley remains clearly identifiable over a wide range of $\mathrm{[Mg/Si]}$. This difference likely reflects both the more complex nucleosynthetic origin and Galactic evolution of carbon relative to the $\alpha$-elements Mg and Si \citep{2013pss5.book...21N}, as well as a weaker correspondence between stellar C and O abundances and planetary compositions. In contrast, refractory rock-forming elements such as Mg and Si are expected to be more directly inherited by planets, consistent with recent measurements showing stellar-like Mg/Si and Fe/Mg ratios in giant-planet atmospheres \citep{2025arXiv251210904S}. Together, these results suggest that elemental ratios closely linked to planetary interior composition serve as robust tracers of the physical processes shaping the radius valley.

\begin{figure*}
    \centering
    \includegraphics[width=\linewidth]{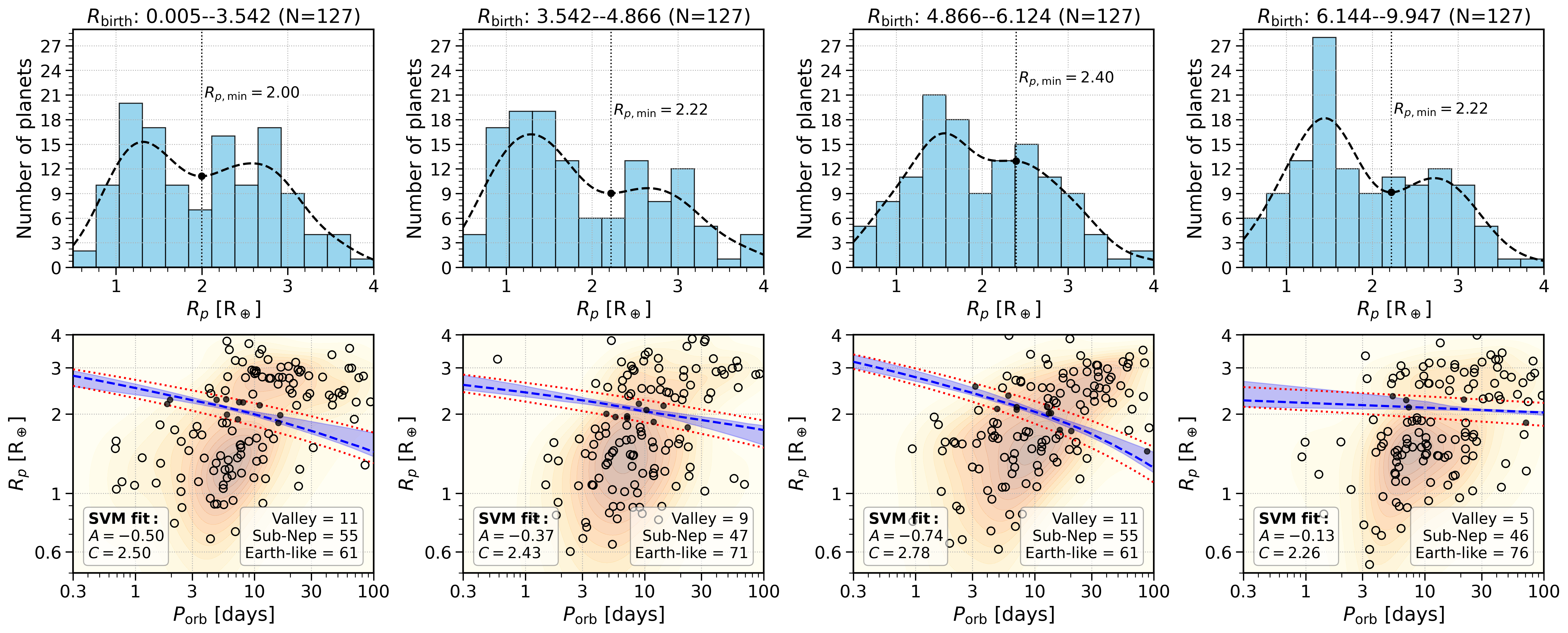}
    \caption{
    Same as Figure~\ref{fig:withage}, but binned by stellar birth radius ($R_{birth}$).}
    \label{fig:withrb}
\end{figure*}

\begin{figure}
    \centering
    \includegraphics[width=\linewidth]{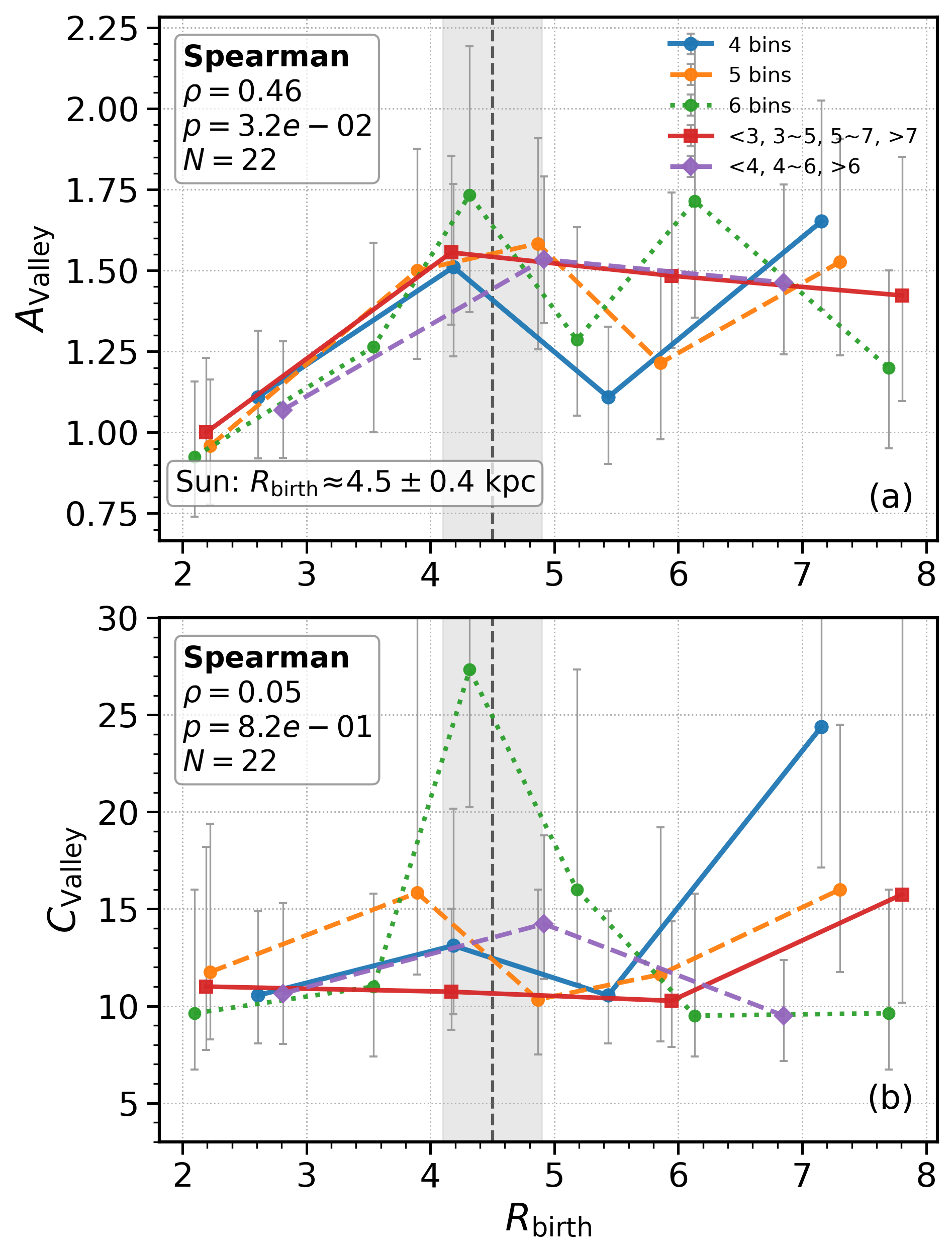}
    \caption{
    Evolution of planet-number ratios with stellar birth radius $R_{\rm birth}$. Symbols and line styles are the same as in Figure~\ref{fig:parawithage}. The shaded vertical band marks the estimated birth radius of the Sun, $R_{\rm birth,\odot} \simeq 4.5 \pm 0.4$~kpc \citep{2024MNRAS.535..392L}.}
    \label{fig:parawithrb}
\end{figure}
\subsection{The Radius Valley across Stellar Birth Radius}

The stellar birth radius, $R_{\rm birth}$, provides a physically motivated link between planet populations and the Galactic environment in which their host stars formed. Unlike the present-day Galactocentric distance, $R_{\rm birth}$ traces the chemical and structural conditions of the natal interstellar medium. Because both planet formation and atmospheric evolution depend on these environmental properties, the variation of the radius valley with $R_{\rm birth}$ offers a direct probe of Galactic environmental effects on close-in planets.

Figures~\ref{fig:withrb} show that the morphology of the radius valley varies systematically with $R_{\rm birth}$. Stars born in the inner disk ($R_{\rm birth} \lesssim 6.1$~kpc) exhibit a clear deficit near $R_p \sim 2\,R_\oplus$ and a strong period dependence of the valley location, with SVM slopes ranging from $-0.74$ to $-0.37$, indicating a well-defined bimodal structure. In contrast, the outermost bin ($R_{\rm birth}=6.1$--9.9~kpc), corresponding to stars formed near the local disk, shows a much weaker deficit around $2\,R_\oplus$ and a substantially flatter valley slope ($-0.13$), although it contains the largest number of Earth-like planets (76 planets, 59\%). Figure~\ref{fig:parawithrb} shows the dependence of $A_{\rm valley}$ and $C_{\rm valley}$ on stellar birth radius. The results are broadly consistent across different binning schemes and reveal a non-monotonic trend. Both parameters are lowest at the smallest $R_{\rm birth}$, increase toward intermediate birth radius, reach a maximum around $R_{\rm birth} \sim 4$--5~kpc, and then decline toward $R_{\rm birth} \sim 5$--6~kpc before rising again at larger birth radius. 

The systematic variation of the valley morphology with $R_{\rm birth}$ indicates that the radius valley retains a memory of the Galactic environment in which planetary systems formed rather than being determined solely by present-day stellar properties. Although all systems are observed near the Solar neighborhood, stars born at different radius are expected to have undergone radial migration, bringing planetary systems formed under distinct disk conditions into the local sample. Systems originating in the inner Galaxy, where gas and solid surface densities were higher, likely experienced more efficient core growth and more uniform formation conditions, producing a well-separated bimodal planet population and a clearly defined valley. In contrast, stars associated with the local disk span a broad range of birth radius, approximately 6--10 kpc. This extended radial coverage encompasses a diverse set of formation environments and radial migration histories, which likely give rise to a wider dispersion in planetary core masses and atmospheric mass-loss pathways. As a consequence, the resulting radius valley is more diffuse and exhibits a shallower, less well-defined signature. Interestingly, recent dynamical reconstructions suggest that the Sun was born at $R_{\rm birth} \simeq 4.5 \pm 0.4$~kpc \citep{2024MNRAS.535..392L}, close to the first peak of both $A_{\rm valley}$ and $C_{\rm valley}$ identified in Figure~\ref{fig:parawithrb}. This correspondence suggests that the Solar System formed in a Galactic environment where the small-planet population had already developed a clear bimodal structure, consistent with efficient atmospheric evolution and a high fraction of Earth-sized planets. Taken together, these results suggest that the morphology of the radius valley is primarily set by formation environment and subsequent dynamical redistribution within the Galactic disk, linking small-planet demographics to the chemical and structural evolution of the Milky Way.

\section{Summary} \label{sec:con}

We present a systematic investigation of how the exoplanet radius valley depends on multi-dimensional host-star properties, using a homogeneous sample of 769 planets orbiting 558 Kepler host stars cross-matched with LAMOST DR9 DD-Payne and Gaia DR3. By re-deriving stellar luminosities and radius, recomputing planet radius from transit depths, and constructing parallel subsamples with reliable ages, chromospheric activity ($\log R^+_{\rm HK}$), detailed elemental abundances, and inferred stellar birth radius, we quantified the valley morphology via an SVM-defined locus and two number-ratio diagnostics, $A_{\rm valley}$ and $C_{\rm valley}$. Our main results are as follows.

\begin{enumerate}
\item \textit{Age dependence.} The radius valley is weak or absent in the youngest systems ($\lesssim 3$~Gyr) and becomes more distinct on gigayear timescales. Its evolution is clearly non-monotonic, with intermediate-age bins showing an apparent weakening that is naturally explained by population mixing rather than a reversal of planetary evolution. This delayed emergence and Gyr-scale evolution is more consistent with gradual atmospheric loss as planets cool and contract than with a purely early-time origin.

\item \textit{Magnetic activity.} The valley morphology varies non-monotonically with $\log R^+_{\rm HK}$. A clear valley is present among magnetically quiet stars, while the most active subsamples do not show a systematically deeper depletion. This weak and non-monotonic dependence of valley contrast on activity suggests that high-energy irradiation is not the sole control parameter of the valley depth, although activity may still modulate the detailed geometry of the valley locus in the $R_p$--$P_{\rm orb}$ plane.

\item \textit{Stellar chemistry.} The valley contrast shows a non-monotonic dependence on metallicity: it is strongest in metal-poor stars, weakens near solar metallicity, and partially strengthens again at the highest metallicities. In addition, the valley displays sensitivity to refractory element ratios such as [Mg/Si]. These results indicate that host-star composition primarily influences the valley through the underlying distribution of core masses and interior structure rather than by directly shifting the valley locus.

\item \textit{Birth-radius imprint.} The radius valley varies systematically with stellar birth radius $R_{\rm birth}$, showing a non-monotonic behavior in both $A_{\rm valley}$ and $C_{\rm valley}$. Systems formed in the inner disk exhibit a clearer bimodal structure and a stronger valley, whereas those associated with the local disk show a weaker and more diffuse signature, likely reflecting population mixing due to radial migration. Notably, the estimated solar birth radius, $R_{\rm birth,\odot}\simeq 4.5\pm0.4$~kpc, lies near the peak of the valley contrast, suggesting that the Solar System formed in a Galactic environment favorable for efficient small-planet formation and long-term atmospheric evolution.

\end{enumerate}

Taken together, the combined constraints from stellar age, magnetic activity, detailed chemistry, and birth radius favor a picture in which the radius valley is primarily shaped by formation-linked core properties and long-timescale planetary cooling---as expected in core-powered mass-loss scenarios---with stellar irradiation playing a secondary role in modulating the detailed valley locus. More broadly, our results demonstrate that the radius valley is not only a diagnostic of planetary atmospheric evolution, but also a tracer of how Galactic environment and disk evolution imprint on the demographics of close-in small planets. Future tests using larger and more precise samples (e.g., Gaia-enabled spectroscopy combined with TESS/PLATO planets and improved dynamical birth-radius inferences) will enable tighter separation of planetary-evolution physics from Galactic population effects and will clarify the extent to which the Solar System is representative of its birth environment.

\begin{acknowledgments}
This work is based on data acquired through the Guoshoujing Telescope. Guoshoujing Telescope (the Large Sky Area Multi-Object Fiber Spectroscopic Telescope; LAMOST) is a National Major Scientific Project built by the Chinese Academy of Sciences. Funding for the project has been provided by the National Development and Reform Commission. LAMOST is operated and managed by the National Astronomical Observatories, Chinese Academy of Sciences. This work has made use of data from the European Space Agency (ESA) mission Gaia (\url{https://www.cosmos.esa.int/gaia}), processed by the Gaia Data Processing and Analysis Consortium (DPAC, \url{https://www.cosmos.esa.int/web/gaia/dpac/consortium}). Funding for the DPAC has been provided by national institutions, in particular the institutions participating in the Gaia Multilateral Agreement. We acknowledge the entire Kepler team and everyone involved in the Kepler mission. Funding for the Kepler mission is provided by NASA's Science Mission Directorate.
This work was supported by the National Natural Science Foundation of China (Grants No.12403037, Grants No.12503025).
\end{acknowledgments}

\clearpage
\bibliography{Biblio}{}
\bibliographystyle{aasjournalv7}



\end{document}